\documentclass{openjournal}[revtex] 

\pdfoutput=1 

\usepackage[T1]{fontenc} 
\usepackage{graphicx}
\usepackage{bm}
\usepackage{color}
\usepackage{makecell}
\usepackage{afterpage}
\usepackage[dvipsnames]{xcolor}
\usepackage{multirow}

\usepackage{subfigure}
\usepackage{booktabs}
\usepackage[normalem]{ulem}		
\usepackage{pgfgantt}
\usepackage{cancel}
\usepackage{chngcntr}
\usepackage{amsmath}

\usepackage[breaklinks,colorlinks,citecolor=blue,linkcolor=blue,urlcolor=blue]{hyperref}

\newcommand{\dd}{{\rm{d}}}

\newcommand{\fett}[1]{{\boldsymbol{#1}}}

\newcommand{\be}{\begin{equation}}
\newcommand{\ee}{\end{equation}}

\def \HH {\mathcal{H}}

\newcommand{\Hc}{\mathcal{H}}

\begin{document}
\hfill {\scriptsize RBI-ThPhys-2025-01}

\title{Angular bispectrum of matter number counts in cosmic structures}
\author{Thomas Montandon}
\affiliation{Laboratoire Univers et Particules de Montpellier, Universit\'e de Montpellier/CNRS, place E. Bataillon, cc072, 34095 Montpellier, France}


\author{Enea Di Dio}
\affiliation{Universit\'e de Gen\`eve, D\'epartement de Physique Th\'eorique and Centre for Astroparticle Physics,
24 quai Ernest-Ansermet, CH-1211 Gen\`eve 4, Switzerland}

\author{Cornelius Rampf}
\affiliation{Division of Theoretical Physics, Ru\dj er Bo\v{s}kovi\'c Institute, Bijeni\v{c}ka cesta 54, 10000 Zagreb, Croatia}

\affiliation{Department of Astrophysics, University of Vienna, T\"urkenschanzstraße 17, 1180 Vienna, Austria}

\affiliation{Department of Mathematics, University of Vienna, Oskar-Morgenstern-Platz 1, 1090 Vienna, Austria}

\author{Julian Adamek}
\affiliation{Institut f\"ur Astrophysik, Universit\"at Z\"urich, Winterthurerstrasse 190, 8057
Z\"urich, Switzerland}

\begin{abstract} 
    The bispectrum of galaxy number counts is a key probe of large-scale structure, offering insights into the initial conditions of the Universe, the nature of gravity, and cosmological parameters. In this work, we present the first full-sky computation of the angular bispectrum in second-order perturbation theory without invoking the Limber approximation, and formulated for finite redshift bins via window functions, i.e., using tomographic spherical harmonics. To our knowledge, even the Newtonian part within this setup is novel. Building on this, we also include, up to second order in perturbation theory, the dynamical general relativistic and radiation effects, together with the leading relativistic projection effects. For simplicity, we neglect tracer bias and line-of-sight integrated contributions, however note that in particular the former can be straightforwardly incorporated within our framework. We evaluate the bispectrum contributions for two redshift bins, $1.75 \leq z \leq 2.25$ and $0.55 \leq z \leq 0.65$, and compare our theoretical prediction against relativistic light-cone simulations, with line-of-sight integral effects removed so as to enable direct consistency checks. As expected, we find that the Newtonian contributions are typically one or more orders of magnitudes larger than the relativistic signal across the entire spectrum for both redshifts. At $z=2$, we find that projection and dynamical relativistic effects have comparable amplitudes on large scales; somewhat unexpectedly, however, radiation effects dominate the relativistic signal in the squeezed limit. At $z=0.6$, the expected hierarchy is recovered, though dynamical corrections (from nonlinear evolution) remain non-negligible --- only a factor of 2–3 smaller than projection effects. Our theoretical results agree fairly well with simulation measurements for the total bispectrum. To facilitate future applications and reproducibility, we make the corresponding code publicly available.
\end{abstract}

\section{Introduction} 
The two-point correlation function, or its Fourier counterpart -- the power spectrum -- remains the primary source of cosmological information used to constrain the standard $\Lambda$CDM model and its extensions. In the linear regime, where fluctuations are small, the two-point statistic dominates in signal-to-noise and is significantly easier to measure than higher-order correlators. In the weakly nonlinear regime in contrast, model uncertainties grow, and shot noise further limits the extraction of higher-order statistics. The most stringent cosmological constraints to date stem from observations of the cosmic microwave background (CMB). The Planck collaboration measurements~\citep{Planck:2019kim, Planck:2019evm} show no deviations from statistical homogeneity and isotropy, and are consistent with Gaussian, adiabatic initial conditions. As a result, the statistical properties of large-scale cosmological fields, such as the matter density and velocity, are now tightly constrained by the power spectrum alone \cite[see][]{Planck:2018vyg, ACT:2020goa, DES:2022urg, eBOSS:2021poy, DESI:2024mwx, DESI:2024hhd}.

In the coming years, an influx of high-precision data from upcoming photometric and spectroscopic surveys -- including Euclid~\citep{EUCLID:2011zbd}, the Vera Rubin Observatory~\citep{LSSTScience:2009jmu}, SPHEREx~\citep{SPHEREx:2014bgr}, the Nancy Grace Roman Space Telescope~\citep{Spergel:2015sza}, and the Chinese Space Station Telescope~\citep{Gong:2019yxt} -- will further enhance our ability to test $\Lambda$CDM. Further ahead, the Square Kilometre Array Observatory (SKAO) will probe the large-scale structure via $21$-cm neutral hydrogen emission~\cite[see e.g.][]{SKA:2018ckk, Bull:2015lja}. This unprecedented dataset offers the opportunity not only to tighten existing constraints but also to uncover subtle effects of new physics -- whether through cosmic tensions such as the $H_0$ discrepancy~\citep{Riess:2016jrr, Poulin:2023lkg, DiValentino:2021izs}, signatures of modified gravity~\citep{Beutler:2020evf}, nature of dark energy from which DESI already provided strong hints on a time evolution \citep{DESI:2024mwx}, or the physics of dark matter or early universe through primordial non-Gaussianity~\citep[PNG;][]{Sailer:2021yzm}. 

Yet, the late-time large-scale structure (LSS) evolves nonlinearly, leading to non-Gaussian features that are not captured by the two-point function alone. For a more complete description of the statistical properties of the matter distribution, higher-order statistics are necessary. In particular, the three-point correlation function and its Fourier transform -- the bispectrum -- play a crucial role in probing Fourier phase correlations and mode couplings induced by nonlinear gravitational evolution or PNG.

The CMB bispectrum has been investigated with both WMAP~\citep{Fergusson:2010dm} and Planck~\citep{Planck:2019kim}, though without significant PNG detection. In contrast, galaxy surveys have already yielded bispectrum measurements. The first detection came from the IRAS Point Source Catalog Redshift Survey~\citep[PSCz;][]{Scoccimarro:1997st}, with later surveys like BOSS~\citep{Gil-Marin:2014sta} refining the signal~\citep{Gil-Marin:2016wya, Cabass:2022wjy, DAmico:2022gki, Ivanov:2023qzb}. 
However, these analyses have so far relied on the local flat-sky approximation~\citep{Scoccimarro:2015bla} --- where the line of sight varies throughout the survey but each pair or triplet of galaxies is assumed to have parallel lines of sight when computing correlations --- the Limber approximation \cite[see][]{Limber:1954zz, Kaiser:1996tp, LoVerde:2008re} and Newtonian approximation where only the Newtonian gravitational dynamics (including RSDs) are considered, see~\citet{Bernardeau:2001qr}.
The next generation of galaxy surveys -- especially those combining spectroscopic and photometric observations -- will offer richer data and call for more sophisticated bispectrum analyses.

In spectroscopic surveys such as the one of Euclid or DESI,
precise redshift measurements with an error $\sigma_z = 10^{-3}(1+z)$ \citep{Euclid:2024yrr} enable the reconstruction of the three-dimensional galaxy distribution. Euclid, for instance, will collect spectroscopic data for $\sim$30 million galaxies over one-third of the sky \citep{Euclid:2024yrr} -- an order of magnitude more than BOSS and nearly the same number as that of DESI~\citep{DESI:2025fxa}. Several frameworks have been developed to extract the bispectrum in redshift space, including Legendre polynomial decompositions like the Yamamoto–Scoccimarro formalism~\citep{Scoccimarro:2015bla} and tripolar expansions~\citep{Sugiyama:2018yzo}, supported by theoretical predictions~\cite[see e.g.][]{Scoccimarro:1999ed, Nan:2017oaq, deWeerd:2019cae}. However, as the survey volume increases, wide-angle effects become important, and the flat-sky approximation breaks down. These effects can be partially modeled within Fourier-space approaches~\citep{Castorina:2017inr, Beutler:2018vpe, Castorina:2021xzs, Noorikuhani:2022bwc, Pardede:2023ddq, Foglieni:2023xca}, but their inclusion in bispectrum analyses remains challenging. Additionally, when correlating points at unequal redshifts, radial evolution effects stemming from the background should be taken into account~\citep{Bonvin:2013ogt, Beutler:2020evf, Addis:2024zhw}.
One solution to overcome these effects within our framework is to use directly angular statistics integrated over redshift bins.  Additionally, on large scales, Newtonian approximation may become insufficient, motivating a check for relativistic effects as attempted in this work.

One well-suited approach --- applicable to angular statistics --- for spectroscopic surveys is to use the spherical Fourier-Bessel (SFB) decomposition first introduced by~\citet{1973ApJ...185..413P}. \citet{Bertacca:2017dzm} derived the bispectrum expression, while \citet{Benabou:2023ldb} performed the numerical evaluation of the dominant terms. 
Euclid will also carry out a wide-area photometric survey over the same sky region, detecting about 1.5 billion galaxies \citep{Euclid:2024yrr}. Due to limited redshift resolution ($\sigma_z = 0.05 (1+z)$), the data will be divided into a handful of tomographic bins. In such cases, the SFB decomposition is no longer optimal, as the poor radial resolution prevents a faithful recovery of the full 3D modes. Instead, these projected fields require angular statistics tailored to discrete redshift bins. This approach is sometimes referred as ``tomographic spherical harmonics'' (TSH), the approach that we consider in this work. Such angular statistics estimators are already well developed in the context of the CMB~\cite[see e.g.][]{Planck:2019kim} and can be adapted to galaxy clustering. For instance, \citet{Montandon:2022ulz} used the binned bispectrum estimator of \citet{Bucher:2015ura}. However, the theoretical prediction of the angular statistics of the number count involves slowly converging integrals over spherical Bessel functions. Mathematical techniques and efficient numerical methods for evaluating such integrals (e.g. efficient evaluation of hyper-geometric functions, FFTLog decompositions and suitable integration by part) has been studied in great details and are being applied to the power spectrum by~\citet{Assassi:2017lea} and \citet{Simonovic:2017mhp} \citep[see also][]{GrasshornGebhardt:2017tbv, Fang:2019xat}. The focus of the present work is to use the framework of~\citet{Assassi:2017lea} to compute the angular bispectrum of the matter number count.

\cite{DiDio:2014lka, DiDio:2015bua} first derived the angular bispectrum of galaxy number counts under simplifying assumptions such as infinitesimal redshift bins, Newtonian dynamics, and the Limber approximation. 
\cite{DiDio:2018unb} then incorporated finite redshift binning and redshift-space distortions (RSD) but still relied on the Limber approximation to reduce computational cost. An accompanying code, \href{https://gitlab.com/montanari/byspectrum}{\texttt{byspectrum}}, was released but was limited to the single density term and infinitesimal redshift bins. \cite{Assassi:2017lea} also applied their efficient mathematical framework to evaluate the Newtonian density term of the matter number count bispectrum and provides a \texttt{Mathematica} notebook.
Together with this work, we provide the publicly available \href{https://github.com/TomaMTD/ang_bispec}{\texttt{ang\_bispec}} code that essentially extends the computations of~\cite{Assassi:2017lea} to all the dominant Newtonian terms, i.e., density, RSD and quadratic terms, and to relativistic effects. 

In the meantime, \citet{Adamek:2021rot} and \cite{Montandon:2022ulz} developed relativistic $N$-body simulation pipelines that also go beyond these limitations, producing fully relativistic light-cone observables and enabling direct comparison with Newtonian approximations. These simulations allow the measurement of the angular bispectrum of matter number counts with relativistic dynamics, analysed using the binned estimator of~\cite{Bucher:2015ura}. 
From these simulations, the total matter bispectrum on the light cone was measured with solid significance, however we note that extracting subtle (relativistic) contributions from simulations can be a delicate matter. Also, extensions to simulations of biased tracers are still on an early stage due to computational cost limitation~\cite[e.g.][]{Montandon:2024mku}.

In this work, we attempt to rectify some of these shortcomings; in particular, 
\begin{enumerate}
    \item  we evaluate the theoretical tree-level angular bispectrum in perturbation theory with finite redshift bin width without resorting to the Limber approximation. To our knowledge, this `Newtonian' calculation --- stemming from the Newtonian matter density and RSDs --- has not been reported previously in the literature within the present setup.
    \item Up to second order, we include all ``dynamical'' GR effects that stem from nonlinear evolution \citep{Matarrese:1997ay, Bartolo:2010rw, Uggla:2013kya, Bruni:2013qta, Villa:2015ppa}, and furthermore include radiation corrections~\citep{Bartolo:2005kv, Fitzpatrick:2009ci, Bartolo:2006fj, Pitrou:2010sn, Huang:2012ub, Su:2012gt, Huang:2013qua, Pettinari:2013he, Pettinari:2014vja,Tram:2016cpy}.
    \item We compare our novel theoretical predictions for both the Newtonian and GR contributions against the relativistic $N$-body simulation results of~\cite{Montandon:2022ulz}, for which we employ binning and smoothing \cite[see][]{Bucher:2015ura}:
    binning implies averaging the measured bispectrum for fixed triangle configurations over multipole windows, while the smoothing operation (with a Gaussian), applied to the binned bispectrum, increases the signal-to-noise ratio even more (see Sec.\ \ref{sec:Comparison with simulation} for details). 
    \item We also include all non-integrated relativistic projection effects up to order $\mathcal{H}/k$~\citep{Yoo:2009au,Yoo:2010ni,Challinor:2011bk,Bonvin:2011bg,Jeong:2011as,Ginat:2021nww, Durrer:2016jzq,DiDio:2016ykq}, also those appearing at second order \citep{DiDio:2014lka,Clarkson:2018dwn,DiDio:2018zmk,DiDio:2020jvo},  allowing us to compare the spatio-temporal footprint of these effects against generic GR effects (see point 2) within a unified and controlled setup.  
    Here, ``up to order ${\cal H}/k$'' refers to a (standard) weak-field hierarchy at the field level, where Newtonian density terms scale as $({\cal H}/k)^0$, velocity terms as ${\cal H}/k$, and metric perturbations as $({\cal H}/k)^2$. 
\end{enumerate}
We acknowledge that our computation is not exhaustive; specifically,
\begin{itemize}
    \item we ignore integrated projection effects such as lensing and Integrated Sachs–Wolfe. The computation of these contributions requires a more involved formalism with additional line-of-sight integrals, which we leave for future work. 
    These terms can however become important. \cite{DiDio:2016gpd}
showed that lensing terms can become dominant for cross-correlations between well-separated redshift bins, while for auto-bispectra (all three redshifts equal), they can reach $10\%$ of the total amplitude for wide redshift bins. For simplicity, we therefore focus on the auto-bispectrum regime and enable comparisons by removing these integrated effects from the relativistic simulation results. 
Future work incorporating lensing terms would enable cross-bin-correlation studies and measurement of lensing signals through redshift separation techniques.
    \item We ignore all tracer bias effects, as no relativistic galaxy simulations currently exist to enable a meaningful comparison with our theoretical predictions. Nonetheless, we remark that effects such as linear and nonlinear galaxy bias, magnification bias, and time-evolution bias can be directly incorporated within our formalism.
    \item We retain next-to-leading projection effects $({\cal H}/k)^2$ at first order, but omit them at second order. Although these could, in principle, be implemented within our framework, the process is tedious and the resulting contributions are expected to be vastly suppressed compared to the leading projection terms we do include. 
    We also note that, while different derivations of these next-to-leading projection effects exist~\citep{Yoo:2014sfa,Bertacca:2014dra,DiDio:2014lka,Magi:2022nfy}, they did not yet converge to a consensus due to their huge complexity.
\end{itemize}
In summary, we include in total 7 linear and 14 second-order terms in the theoretical computation, summarised and described in Section~\ref{sec:Number Counts}; see specifically Eqs.\,\eqref{eq:1stNB} and~\eqref{eq:2ndNB} for the first and second-order contributions, respectively.
In Section~\ref{sec:angular bispectrum},  we present the angular bispectrum using tomographic bin statistics in a form that is convenient for numerical evaluation.
In Section~\ref{sec:Theoretical results}, we group all considered terms into eight bispectrum contributions, organised into physically motivated categories (Eq.\,\ref{eq:b_split}), and provide an in-depth analysis. Section~\ref{sec:Comparison with simulation} compares our predictions with relativistic simulations performed by \cite{Montandon:2022ulz}, and conclusions follow in Section~\ref{sec:conclusion}.

\section{Number Counts}\label{sec:Number Counts}
We start by defining the number count perturbation as the fluctuation of the number of discrete sources $N(\boldsymbol n, z)$ in an angular direction on the sky $\boldsymbol n$ and at observed redshift $z$,
\be
\Delta (\boldsymbol n, z) = \frac{N (\boldsymbol n, z) - \left<N\right> (z)}{\left<N\right> (z)}\,,
\ee
where $\langle \ldots \rangle$ denotes the ensemble average.
Given that the number count perturbation $\Delta (\boldsymbol n, z)$ is an observable, and therefore gauge-invariant, we can compute it in any gauge. We choose to work in Newtonian gauge where the perturbed FLRW line element is written as 
\begin{equation}
ds^2= a^2 \left[ -\left( 1+2 \psi \right) d\tau^2 + \left( 1- 2 \phi \right) d{\boldsymbol{x}}^2 \right]\,,
\end{equation}
where $a$ denotes the cosmic scale factor.
The Newtonian gauge is a restriction of Poisson gauge, where one neglects vector and tensor perturbations of the metric and only keeps the two gravitational potentials $\psi$ and $\phi$. Those potentials are assumed to be in a weak-field regime in cosmology, and we shall consider the leading weak-field order only. This means that we neglect terms that are quadratic in $\phi$, $\psi$, and we have expanded the above expression for the line element accordingly.
In standard perturbation theory, we expand the number count perturbation of the density as 
\be\label{eq:SPT}
  \Delta = \sum_{i\ge 1} \Delta_i \,,
\ee
and similarly for the velocity and metric perturbations.

On top of the standard perturbation theory expansion \eqref{eq:SPT}, we organise terms according to the weak-field expansion parameter $\mathcal H/k$. The ``Newtonian terms'' (including RSDs) refer to the leading order $(\mathcal H/k)^0$ that dominate at small scales while relativistic terms refers to higher powers of $\mathcal H/k$.

\subsection{First order}\label{sec:FO}
 The expression of the linear number count perturbation is well known \cite[see][]{Yoo:2009au,Yoo:2010ni,Challinor:2011bk,Bonvin:2011bg,Jeong:2011as}, and (by neglecting integrated terms such as lensing convergence, integrated Sachs-Wolfe and time-delay effects) leads to\footnote{We remark again that we are considering the matter number counts, therefore we do not have galaxy bias nor evolution bias. We also neglect magnification bias that would affect flux-limited galaxy surveys. We also omit the terms evaluated at the observer position. However, these do not contribute to the tree-level angular bispectrum from $\ell > 2$ as shown by~\cite{DiDio:2014lka}. All the perturbations are expressed in Newtonian gauge.}
\be\label{eq:1stNB}
    \Delta_1 = \Delta_1^{\rm N} + P^{\rm R}_1 + D_1^{\rm GR}\,,
\ee
where we have split the number count perturbation into a Newtonian $(\rm N)$ part $\Delta_1^{\rm N} \propto (\mathcal H / k)^0$, a projection part, $P_1^{\rm R}$, whose leading term is $\propto (\mathcal H / k)$ and a dynamical GR part $D_1^{\rm GR} \propto (\mathcal H / k)^2$ which read
\begin{align}
    \Delta_1^{\rm N} &= \delta^{\rm N}_{1} - \mathcal H^{-1}\partial_r^2 v_1 \label{eq:delta1_n}\,,\\ 
    P_1^{\rm R} &=  - \mathcal R \partial_r v_1 - 2 \phi_1+ (\mathcal R+1 )\psi_1 + \mathcal H^{-1} \dot \phi_1\label{eq:delta1_p}\,,\\
    D_1^{\rm GR} &= \delta^{\rm GR}_{1} \label{eq:delta1_gr}\,,
\end{align}
where a dot denotes the partial derivative with respect the conformal time $\tau$, $r$ is the comoving distance, $\mathcal{H}=\dot a/a$ is the conformal Hubble parameter, $v_1$ the first-order velocity potential related to the peculiar velocity through $\boldsymbol{\nabla} v_1 = \boldsymbol{v}$, $\delta_1$, $\phi_1$ and $\psi_1$ are the first-order density contrast and potentials, and
\be
    \mathcal{R}=\left(\frac{\mathcal{\dot{H}}}{\mathcal H^2} + \frac{2}{r \mathcal H} \right) \, .
\ee
The first term of Eq.\,\eqref{eq:1stNB} represents the standard Newtonian perturbation theory, i.e.~density plus redshift-space distortions, see Eq.\,\eqref{eq:delta1_n}. The remaining terms are due to relativistic effects. These effects are split into two different terms, one accounting for projection effects defined in Eq.\,\eqref{eq:delta1_p} and another one for dynamical GR effects defined in Eq.\,\eqref{eq:delta1_gr}. Let us stress that Eq.\,\eqref{eq:1stNB} is derived only under the assumption that photons travel along light-like geodesics and that matter motion is described by the Euler equation (i.e.~galaxies follow time-like geodesics). Among the linear projection effects we have a Doppler term $\partial_r v_1$ which in the weak-field approximation is suppressed by a factor $\mathcal H/k$ with respect to density fluctuations, and terms which are directly proportional to the metric perturbations (as Sachs--Wolfe terms) which are suppressed by a factor $\left( \mathcal H /k \right)^2$ with respect to Newtonian perturbation theory. 
These terms are included for completeness; however, since we consider projection effects only up to $\mathcal H/k$ at second order, we focus solely on the leading-order contributions in the following.

To understand the last relation \eqref{eq:delta1_gr}, we stress that all the perturbations sourcing the linear number counts in Eq.\,\eqref{eq:1stNB} are related to each other through the theory of gravity. In GR, in the absence of large-scale anisotropic stress, i.e. neglecting radiation, we can express all perturbations in terms of the present-day linear gravitational potential $\phi_0$, as $\phi = \phi_0 \times D/a$, where $D$ denotes the linear growth factor. Following the conventions of \cite{Villa:2015ppa}, we use a $\Lambda$CDM cosmology without radiation such that $\phi_1=\psi_1$, the linear density contrast
\be\label{eq:density2pot}
     \delta_1(\tau, \boldsymbol k) = - \frac{2}{3} \frac{D k^2}{\mathcal H_0^2 \Omega_{m0}} \left( 1 + 3f \frac{\mathcal H^2}{k^2} \right) \phi_0(\boldsymbol k) 
\ee
(the first [second] term on the r.h.s.\ equals to $\delta_1^{\rm N}$ [$\delta_1^{\rm GR}$]),
and the velocity potential
\be\label{eq:velocity2pot}
      v_1(\tau, \boldsymbol k) = -\frac{2 D \mathcal H f}{3 \mathcal H_0^2 \Omega_{m0}} \phi_0(\boldsymbol k) \,,
\ee
where $f = \dd \ln D / \dd \ln a$ is the linear growth rate.
Furthermore,
\begin{equation}\label{eq:dpot}
      \dot \phi_1(\tau, \boldsymbol k) = \frac{\mathcal H D}{a} (f-1) \phi_0(\boldsymbol k)\,.
\end{equation} 
From Eqs.\,\eqref{eq:density2pot}, \eqref{eq:velocity2pot}, and \eqref{eq:dpot},  we see that, at the linear level, there is only a single GR correction that affects the density, which appears in Eq.\,\eqref{eq:delta1_gr}.

\subsection{Second order}\label{sec:SO}

Next we consider the second-order number count \cite[see][for derivations]{Yoo:2014sfa,Bertacca:2014dra,DiDio:2014lka,Magi:2022nfy}.
\cite{DiDio:2018zmk,Clarkson:2018dwn,DiDio:2020jvo} showed that these derivations agree at least up to the first order in the weak-field expansion $\mathcal H/k$. We therefore limit our analysis to these terms, and furthermore neglect integrated terms consistently with the linear-order expression~\eqref{eq:1stNB}; see the end of the introduction for related explanations and justifications. Furthermore, we include the radiative correction at second order by following \cite{Tram:2016cpy}: Specifically, \cite{Tram:2016cpy} employs the first- and second-order expressions in a base $\Lambda$CDM cosmology without radiation \citep[employing a backscaling approach; see e.g.][]{Michaux:2020yis}, and then adds a second-order radiative correction to this base model, essentially by exploiting a separate universe approach.

The second-order number count reads 
\begin{align}\label{eq:2ndNB}
   \Delta_2 =\; &    \delta_2  - \mathcal{H}^{-1} \partial_r^2 v_2 \nonumber\\
  & + \mathcal{H}^{-2} \left[ \left( \partial_r^2 v_1 \right)^2 + \partial_r v_1 \partial_r^3 v_1\right]  - \mathcal{H}^{-1} \left[  \partial_r v_1 \partial_r \delta_1 + \partial_r^2 v_1  \delta_1\right] \nonumber \\
    &
   -
\mathcal R \partial_r v_2 + \mathcal H^{-1} \left( 1+3\frac{ \dot {\mathcal H}}{\mathcal H^2} + \frac{4}{\mathcal H r}\right) \partial_r v_1 \partial_r^2 v_1 - \mathcal R \partial_r v_1 \delta_1 + \partial_r v_1 \dot \delta_1 + 2 \mathcal H^{-1} v_1^\alpha \partial_\alpha \partial_r v_1 \nonumber  \\
   &- \mathcal H^{-2} \psi_1 \partial^3_r v_1 + \mathcal H^{-1} \psi_1 \partial_r \delta_1 + \mathcal H^{-2} \partial_r v_1 \partial_r^2 \psi_1   \,,  
\end{align}
where $\partial_\alpha$ is the angular derivative over the angle $\alpha=\{\theta,\varphi \}$ 
and we adopt Einstein's index summation convention. 
For convenience, we split the whole expression into different pieces that have different physical origins such that 
\begin{align}\label{eq:split}
    \Delta_2 = \Delta_2^{\rm N} + P_2^{\mathrm R} + D_{2}^{\mathrm {GR}}+ D_{2}^{\mathrm {Rad}} + P_2^{\rm GR}+P_2^{\rm Rad}
\end{align}
where all terms are discussed in the remainder of this section. 
The first term is the Newtonian number count perturbation at order $\mathcal H^0/k^0$, reading 
\begin{equation}\label{eq:Delta2_N}
    \Delta_2^{\mathrm N} = \delta_2^{\mathrm N} - \mathcal{H}^{-1} \partial_r^2 v_2^{\mathrm N} + P_2^{\mathrm N}\,.
\end{equation}
The quantities $\delta_2^{\mathrm N}$ and $v_2^{\mathrm N}$ are the purely second-order Newtonian density and velocity perturbations involving mode couplings, see Appendix \ref{app:Kernel} for an explicit definition. The formal difference to the linear expression is the quadratic Newtonian projection term $P_2^{\mathrm N}$ reading
\begin{equation}\label{eq:QN}
   P_2^{\mathrm N}  = \mathcal{H}^{-2} \left[ \left( \partial_r^2 v_1 \right)^2 + \partial_r v_1 \partial_r^3 v_1\right]  - \mathcal{H}^{-1} \left[  \partial_r v_1 \partial_r \delta^{\mathrm N}_1 + \partial_r^2 v_1  \delta_1^{\mathrm N}\right] \,,
\end{equation}
which, in contrast to $\delta_2^{\mathrm N}$, does not involve any non-trivial coupling kernel. 
The term labelled with superscript $\mathrm R$ in Eq.\,\eqref{eq:split} contains the relativistic leading projection effects 
at order $\mathcal H/k$,
\begin{align}\label{eq:projection}
    P_2^{\mathrm R} =& 
    -\mathcal R \partial_r v_2^{\mathrm N} + \mathcal H^{-1} \left( 1+3\frac{ \dot {\mathcal H}}{\mathcal H^2} + \frac{4}{\mathcal H r}\right) \partial_r v_1 \partial_r^2 v_1 - \mathcal R \partial_r v_1 \delta_1^{\mathrm N} + \partial_r v_1 \dot \delta_1^{\mathrm N} + 2 \mathcal H^{-1} v^a \partial_a \partial_r v_1 \nonumber  \\
    & - \mathcal H^{-2} \psi_1 \partial^3_r v_1 + \mathcal H^{-1} \psi_1 \partial_r \delta_1^{\mathrm N} + \mathcal H^{-2} \partial_r v_1 \partial_r^2 \psi_1 \,.
\end{align}
Note that it is not entirely quadratic because of the first term which involves $\partial_r v_2^{\mathrm N}$.
As explained earlier, the different derivations of the galaxy number counts to second order have never been successfully compared beyond $\mathcal O(\mathcal H/k)$, and at the same time the expressions become challenging to implement such that we only compute these leading terms.

The term $D_2^\mathrm{GR}$ appearing in Eq.\,\eqref{eq:split} is due to general relativistic dynamics in the $\Lambda$CDM base model \citep[see e.g.][]{Villa:2015ppa}, while $D_2^\mathrm{Rad}$ is the radiative correction to this base model \citep[implementing the approach of][]{Tram:2016cpy}; they read
\begin{align}\label{eq:gr}
     D^{\mathrm {GR}}_2 &=    \delta_2^{\mathrm{GR}}  - \mathcal{H}^{-1} \partial_r^2 v_2 ^{\mathrm{GR}}- \mathcal{H}^{-1} \left[  \partial_r v_1 \partial_r \delta^{\mathrm{GR}}_1 + \partial_r^2 v_1  \delta_1^{\mathrm{GR}}\right]  \,,\nonumber\\
     D^{\mathrm {Rad}}_2 &=    \delta_2^{\mathrm {Rad}}  - \mathcal{H}^{-1} \partial_r^2 v_2^{\mathrm {Rad}}\,.
\end{align}
We refer the reader to Appendix~\ref{app:Kernel} for a formal definition of $\delta_2^{\rm GR, Rad}$ and $v_2^{\rm GR, Rad}$. 
See also \cite{Tram:2016cpy} for more details related to the radiation part and its derivation.
At leading order, they are $\mathcal O(\mathcal H^2 /k^2 )$ but, since the terms are well known up to $\mathcal O(\mathcal H^4 /k^4 )$ and easy to calculate, all the dynamical relativistic effects are included in the computation.

The last two terms $P_2^{\rm GR}$ and $P_2^{\rm Rad}$ result from the coupling between relativistic projection effects and relativistic dynamical effects. Although these terms are also of higher order, $\mathcal O(\mathcal H^3/k^3)$, their inclusion in our computation is trivial, hence we choose to retain them;
they read
\begin{align}
    P_2^{\rm GR} &= -\mathcal R \partial_r v_2^{\rm GR} - \mathcal R \partial_r v_1 \delta_1^{\rm GR} + \partial_r v_1 \dot \delta_1^{\rm GR} + \mathcal H^{-1} \psi_1 \partial_r \delta_1^{\rm GR}\,,\\
    P_2^{\rm Rad} &= -\mathcal R \partial_r v_2^{\rm Rad}\,.
\end{align}

In summary, we have explicitly defined the matter number counts, and organized all contributions into Newtonian terms, relativistic projection effects, and dynamical GR and radiation effects. We now proceed to compute the angular bispectrum of the matter number counts.

\section{The angular bispectrum}\label{sec:angular bispectrum}
Here we focus on the projection of the matter number count onto the sphere and provide the relevant expressions for computing the angular bispectrum with tomographic bin statistics. Such statistics involve ``finite redshift bins'' and are well suited for the analysis of photometric surveys (which is the main application of our work); this should be contrasted to the SFB approach of e.g.\ \cite{Benabou:2023ldb} which targets spectroscopic surveys with precise redshift information (see also the introduction for details). We remark that projecting the field-level number count onto the sphere preserves the original ${\cal H}/k$ weak-field hierarchy. 
This follows from the approximate correspondence between the Fourier mode $k$ and the multipole on the sphere $\ell \simeq r k $ where $r$ is the comoving distance of the sphere. In fact, the mode $k \simeq \ell/r$ is in general the dominant contribution to the projection integral -- interestingly, this mode is exactly the one selected by the Limber approximation (which, however, we do not invoke).

From the number count perturbation $\Delta \left( \boldsymbol{n} , z \right)$ introduced in Section~\ref{sec:Number Counts}, we can directly compute the angular bispectrum. Due to statistical isotropy, the angular correlation function depends only on three angles, which translate into three multipoles $\left\{ \ell_1 , \ell_2, \ell_3 \right\}$ in the angular bispectrum. However, the lack of translation invariance along the redshift direction induces dependence on the three separate source redshifts $\left\{ z_1 , z_2 , z_3 \right\}$.
The angular bispectrum is then given by~\citep{Bucher:2015ura}  
\begin{align} \label{eq:angleaverage}
    B^{z_1 , z_2 , z_3 }_{\ell_1\ell_2\ell_3} =  N_{\ell_1 \ell_2 \ell_3} \sum_{m_1m_2m_2} 
    \begin{pmatrix}
\ell_1 & \ell_2 & \ell_3\\
m_1 & m_2 & m_3
\end{pmatrix}
    \left< a^{z_1}_{\ell_1 m_1} a^{z_2}_{\ell_2 m_2} a^{z_3}_{\ell_3 m_3} \right>\,,
\end{align}
where $a_{\ell m}^z$ are the coefficients of the spherical harmonic decomposition of the number count perturbation $\Delta (\boldsymbol n, z)$, namely
\be\label{eq:alm_def}
a_{\ell m } \left( z \right) = \int d \Omega_{\boldsymbol{n}} Y^*_{\ell m } \left( \boldsymbol{n} \right) \Delta \left( \boldsymbol{n}, z \right) \,,
\ee
and
\begin{align}
N_{\ell_1 \ell_2 \ell_3} =
    \begin{pmatrix}
\ell_1 & \ell_2 & \ell_3\\
0 & 0 & 0
\end{pmatrix}
    \sqrt{\frac{(2\ell_1+1)(2\ell_2+1)(2\ell_3+1)}{4\pi}}\,.
\end{align}
In order to compare theoretical predictions with other methods/observations, we include redshift binning. Since we ultimately want to compare our theoretical predictions with the simulations measurements performed by~\cite{Montandon:2022ulz}, which employed top-hat window functions,
we adopt a top-hat function as well (which in our routines can easily be replaced by survey-specific window functions), and furthermore smooth out its sharp edges (see eq.\,\ref{eq:window} below):
This smoothed analytical form is crucial for making the full computation tractable within a reasonable time, as it allows for explicit evaluation of derivatives; for details, see~\cite{Assassi:2017lea}. 

We define the smoothed top-hat function over the redshift range $z_{\rm min} \leq z \leq z_{\rm max}$ as

\be\label{eq:window}
W(r, r_{\rm min}, r_{\rm max}, \sigma) = \left[\frac{1}{2}+\frac{1}{2}\tanh\left(\frac{r - r_{\rm min}}{\sigma}\right)\right] 
\left[\frac{1}{2}-\frac{1}{2}\tanh\left(\frac{r - r_{\rm max}}{\sigma}\right)\right]\,,
\ee
 where $r_{\rm min} = r(z_{\rm min})$ and $r_{\rm max} = r(z_{\rm max})$ define the redshift bin labelled $\hat z$. Throughout this work, we refer to each bin $\hat z$ simply by its redshift range $z_{\rm min} \leq z \leq z_{\rm max}$.

The parameter $\sigma$ sets the smoothness of the bin edges; in the limit $\sigma \rightarrow 0$, the function recovers a pure top-hat shape. Therefore, to ensure a valid comparison with \cite{Montandon:2022ulz}, we remain in the regime $\sigma \ll r_{\rm max} - r_{\rm min}$. To reduce clutter, we omit the explicit labels $r_{\rm min}$, $r_{\rm max}$, and $\sigma$ in the subsequent expressions.

The angular bispectrum for different redshift bins reads
\begin{align}    
B^{\hat z_1\hat z_2\hat z_3}_{\ell_1\ell_2\ell_3} &= \frac{ \int dz_1 dz_2 dz_3 W(r (z_1),\hat z_1)W(r (z_2),\hat z_2)W(r (z_3),\hat z_3) B_{\ell_1\ell_2\ell_3} \left( z_1, z_2 , z_3 \right)}{\int dz_1 W(r(z_1),\hat z_1) \int dz_2 W(r(z_2),\hat z_2) \int dz_3 W(r(z_3),\hat z_3)} \nonumber \\
& = \frac{ \int dr_1 dr_2 dr_3 \tilde W(r_1, \hat z_1 ) \tilde W(r_2, \hat z_2 ) \tilde W(r_3, \hat z_3 ) B_{\ell_1\ell_2\ell_3} \left( r_1,r_2 , r_3 \right) }{\int dr_1 \tilde W(r_1,\hat z_1) \int dr_2 \tilde W(r_2,\hat z_2) \int dr_3 \tilde W(r_3,\hat z_3) } \,,
\end{align}
where we have absorbed the change of integration variable $z \rightarrow r$ in the definition of 
\be
    \tilde W (r, \hat z) = \frac{\mathcal H(r)}{a(r)} W(r, \hat z)  \,.
\ee

From here on, we will consider only the tree-level part to the bispectrum which is composed of correlators of the form $\langle \Delta_2 \Delta_1 \Delta_1 \rangle$, where the $\Delta_{1,2}$'s are respectively given in Eqs.\,\eqref{eq:1stNB} and~\eqref{eq:2ndNB}. Then, owing to the linearity of the correlator, we can write for the tree-level bispectrum
\begin{align}\label{eq:fullbl}
    &B^{\hat z_1 \hat z_2 \hat z_3}_{\ell_1\ell_2\ell_3} = B^{\delta_2}_{\ell_1\ell_2\ell_3} + B^{\partial_r^2 v_2}_{\ell_1\ell_2\ell_3} + B^{(\partial_r^2 v_1)^2}_{\ell_1\ell_2\ell_3}  +B^{\partial_r v_1\partial_r^3 v_1}_{\ell_1\ell_2\ell_3}+B^{\partial_r v_1\partial_r \delta_1}_{\ell_1\ell_2\ell_3}+B^{\partial_r^2 v_1 \delta_1}_{\ell_1\ell_2\ell_3}
    \nonumber\\
    &\quad +B^{\partial_r v_2}_{\ell_1\ell_2\ell_3}
    +B^{\partial_r v_1\partial_r^2 v_1}_{\ell_1\ell_2\ell_3}+B^{\partial_r v_1\delta_1}_{\ell_1\ell_2\ell_3}+B^{\partial_r v_1\dot\delta_1}_{\ell_1\ell_2\ell_3}+B^{v^a \partial_a \partial_r v_1}_{\ell_1\ell_2\ell_3}+B^{\psi_1 \partial_r^3 v_1}_{\ell_1\ell_2\ell_3}+B^{\psi_1 \partial_r \delta_1}_{\ell_1\ell_2\ell_3}+B^{\partial_r v_1\partial_r^2 \psi_1}_{\ell_1\ell_2\ell_3}\,.
\end{align}
The employed short-hand notations on the r.h.s.\ of Eq.\,\eqref{eq:fullbl} comprise implicit definitions for the 14 distinct bispectrum contributions (and we drop their redshift dependence for notational ease), each resembling the 14 individual contributions listed in $\Delta_2$.
Let us remark that Eq.\,\eqref{eq:fullbl} is not a decomposition in the contributions to the bispectrum at different orders in the weak-field approximation because we will also include relativistic effects into the linear parts, which mix the ordering in powers of $\mathcal H/ k$. We can get rid of a normalisation factor in the theoretical expressions by using the reduced angular bispectrum, which reads 
\be
b^{\hat z_1 \hat z_2 \hat z_3}_{\ell_1\ell_2\ell_3} = N_{\ell_1 \ell_2 \ell_3}^{-2} B^{\hat z_1 \hat z_2 \hat z_3}_{\ell_1\ell_2\ell_3}\,.
\ee

In computing the bispectrum, we often encounter correlation between the density field, the velocity potential, and the gravitational potential and their derivatives with the linear number count perturbations given in Eq.\,\eqref{eq:1stNB}.
It is therefore convenient to introduce a generalised power\footnote{We remark that we do not adopt the Limber approximation, whose accuracy in the case of the density bispectrum has been studied by \cite{Assassi:2017lea}, showing large discrepancies compared to the exact solution. This effect is not restricted to the largest scales (low $\ell$-modes), but in the case of correlations across different redshifts it remains non-negligible even beyond the matter--radiation equality scale. Recall that the Limber approximation is based on replacing the spherical Bessel functions with Dirac delta distributions, which leads to
$$
\int_{0}^{\infty} dk\, k^{2} j_{\ell}(k\chi) j_{\ell}(kr) f(k) 
\simeq \frac{\pi}{2r^{2}} f\!\left(\frac{\ell + 1/2}{r} \right) \delta_{\rm D}(\chi - r) .
$$
With this approximation, the integrals appearing in the generalized spectra~\eqref{eq:def_gen_spectrum} can be explicitly carried out analytically.
The Limber approximation accuracy is expected to be even worse for the RSD and Doppler contributions, since the presence of derivatives of the Bessel functions tends to smooth out their peaks. 
}
\begin{align}
\label{eq:def_gen_spectrum}
    C_{\ell}^{(n, m)}(r_1) &= - \frac{2}{\pi} D \left( r_1 \right) {\mathcal N^2} 
    \int dr dk \ \tilde W_{r} D_{r} k^{4+n} P_{\phi_0} \left( k \right) j^{(m)}_{\ell} \left( k r_1 \right) \nonumber\\
    &\times \left[
-k^2\left( 1 + 3 \frac{\Hc_{r}^2}{k^2} f_{r} + \frac{1 - \mathcal R}{a\mathcal N k^2}\right) j_{\ell} \left( k r \right) + k^2 f_{r} j''_{\ell} \left( k r \right) + k\mathcal{R}_{r} f_{r} \Hc_{r}j'_{\ell} \left( k r \right)\right]\,,
\end{align}
where $\mathcal N$ is a normalisation factor defined as
\begin{eqnarray}
    \mathcal N = \frac{2}{3\mathcal H_0^2\Omega_{m0}} \,,
\end{eqnarray} 
and $P_{\phi_0}(k)$ denotes the power spectrum of the gravitational potential evaluated at redshift $z = 0$. We use the notation $j^{(m)}$ to indicate the $m$-th derivative of the spherical Bessel function with respect to its argument. We also use the shorthand notation ``$f_r$'' to denote ``$f(r)$''. 
Here, the index $n$ labels the perturbative order within a weak-field expansion $\propto \left( \Hc/k \right)^n \delta$, while the index $m$ counts the number of radial derivatives. With this formalism, we can write all the spectra as follows 
\begin{equation}\label{eqs:generalized_ps_def}
\begin{aligned}
    C_\ell^{\partial_r v} \left( r_1 \right) &=  f(r_1) \Hc (r_1)  C_\ell^{(-1,1)} \left( r_1 \right) \,,\\
        C_\ell^{\partial_r^2 v} \left( r_1 \right) &=  f(r_1)  C_\ell^{(0,2)} \left( r_1 \right) \,,\\
 C_\ell^{\partial_r^3 v} \left( r_1 \right) &=  \frac{f(r_1)}{\Hc \left(r_1 \right)}  C_\ell^{(1,3)} \left( r_1 \right)\,, \\
  C_\ell^{ \psi} \left( r_1 \right) &= -  \frac{1}{\mathcal{N} a(r_1)}  C_\ell^{(-2,0)} \left( r_1 \right)\,,
  \\
  C_\ell^{\partial_r^2 \psi} \left( r_1 \right) &= -  \frac{1}{\mathcal{N} a(r_1) \Hc^2\left(r_1 \right)}  C_\ell^{(0,2)} \left( r_1 \right)\,,
  \\
    C_\ell^{\delta} \left( r_1 \right) &=    C_\ell^{(0,0)} \left( r_1 \right) + 3 f\left( r_1 \right) \Hc^2 \left( r_1 \right)   C_\ell^{(-2,0)} \left( r_1 \right)\,,\\
    C_\ell^{\partial_r\delta} \left( r_1 \right) &= \frac{1}{\Hc (r_1)}C_\ell^{(1,1)} \left( r_1 \right) + 3 f  \left( r_1 \right) \HH\left( r_1 \right) C_\ell^{(-1,1)} \left( r_1 \right)\,,
   \\
    C_\ell^{ \dot \delta} \left( r_1 \right) &=  f _{r_1} \left[  C^{(0,0)}_\ell \left( r_1 \right)  
    + \left( 3 f_{r_1} \Hc_{r_1}^2 
    + 3 \Hc_{r_1} \frac{\dot f_{r_1}}{f_{r_1}} 
    + 6 \dot \Hc_{r_1}
    \right) C^{(-2,0)}_\ell \left( r_1 \right)
    \right]\,,
\end{aligned}
\end{equation}
where we have used that
\begin{eqnarray}
    \dot \delta = -\mathcal{N}D k^2 \Hc f \left( 1  + 3 f\frac{\Hc^2}{k^2} + 3 \frac{\Hc}{k^2} \frac{\dot f}{f} + 6 \frac{\dot\Hc}{k^2}\right) \phi_0 \,.
\end{eqnarray}
Putting everything together, we have decomposed the redshift-binned angular bispectrum into the 14 tree-level contributions of Eq.\,\eqref{eq:fullbl}. To handle the repeated appearance of the linear terms -- they appear as correlations between the density field, the velocity potential, and the gravitational potential (and their derivatives) -- we introduced the generalized power spectra Eqs.\,\eqref{eq:def_gen_spectrum}–\eqref{eqs:generalized_ps_def}, which provide a unified notation allowing us to write all relevant spectra in a compact way. In the next section we turn to the explicit evaluation of the individual contributions.

\subsection{Pure second-order terms}

We begin by computing the contribution to the bispectrum of terms that appear in standard perturbation theory. Up to second order, these terms fall into two categories: those that arise directly at second order, such as $\delta_2$ and $v_2$, and quadratic terms that are products of linear perturbations. Only the former group involves non-trivial coupling kernel, which we address by integrating over appropriate kernel functions for the density and velocity, denoted respectively as $\mathcal{K}^\delta$ and $\mathcal{K}^v$; see below for implicit definitions. Notably, some contributions to these kernels also arise in Newtonian gravity and are therefore referred to as Newtonian terms.

In standard perturbation theory, the purely second-order terms are the two first terms appearing on the r.h.s.\ of Eq.\,\eqref{eq:fullbl}. These terms involve mode coupling, which manifests as a Fourier convolution involving the kernel~$\mathcal K^{\delta}$,
\begin{equation}\label{eq:def_delta2}
    \delta_2(r_1,\boldsymbol{k}_1) = \mathcal N^2 D^2(r_1) \int \frac{d^3k_2 d^3k_3}{(2\pi)^3} \delta_\mathrm{D}^{(3)}(\boldsymbol k_1 - \boldsymbol{k}_2 -\boldsymbol{k}_3) k_2^2k_3^2\mathcal K^\delta(r_1,k_1,k_2,k_3) \phi_0(\boldsymbol{k}_2) \phi_0(\boldsymbol{k}_3) \,.
\end{equation}
Similarly, the purely second-order contribution to the velocity reads
\begin{equation}\label{eq:def_v2}
    v_2(r_1, \fett k_1) =  \mathcal N^2 D^2(r_1) \frac{\mathcal H_{r_1}}{k_1^2} \int \frac{d^3k_2 d^3k_3}{(2\pi)^3} \delta_\mathrm{D}^{(3)}(\boldsymbol k_1 - \boldsymbol{k}_2 -\boldsymbol{k}_3) k_2^2k_3^2\mathcal K^v(r_1, k_1,k_2,k_3) \phi_0(\boldsymbol{k}_2) \phi_0(\boldsymbol{k}_3)
    \,.
\end{equation}
The expressions of the kernels including relativistic effects in $\Lambda$CDM were calculated by~\cite{Villa:2015ppa}, and with early radiation by~\cite{Tram:2016cpy}. We refer the reader to Appendix~\ref{app:Kernel} for the full expressions. To determine $b^{\delta_2}_{\ell_1\ell_2\ell_3}$ and $b^{\partial_r^2 v_2}_{\ell_1\ell_2\ell_3}$ numerically, we first decompose their kernels, simply denoted as $\mathcal K$, into a sum of powers of $k_1$ such that
\begin{align}\label{eq:kernel_decomp}
    \mathcal K \left(r_1, k_1, k_2 , k_3 \right) 
    &=
    k_1^{-4}  \sum_{mn} f_{mn}^{(-4)}(r_1) k_2^m k_3^n
    +k_1^{-2} \sum_{mn} f_{mn}^{(-2)}(r_1) k_2^m k_3^n \nonumber\\
    &+ \sum_{mn} f_{mn}^{(0)}(r_1) k_2^m k_3^n
    + k_1^2 \sum_{mn} f_{mn}^{(2)}(r_1) k_2^m k_3^n
    +k_1^4 \sum_{mn} f_{mn}^{(4)}(r_1) k_2^m k_3^n\nonumber\\
    &+ \sum_p c_p k_1^{\nu_p-1} \left (k_1^{-4} \sum_{mn} f_{mn}^{(-4,R)}(r_1) k_2^m k_3^n
    +k_1^{-2} \sum_{mn} f_{mn}^{(-2, R)}(r_1) k_2^m k_3^n \right)\,.
\end{align}
The first row of Eq.\,\eqref{eq:kernel_decomp} is of a pure relativistic origin, the second row is a combination of Newtonian and relativistic terms, except for its last term which is purely Newtonian. The last row comes from the radiation term. Unfortunately, the radiation term has a nontrivial dependence in $k_1$. Hence, we use a FFTLog decomposition of the potential transfer function with coefficients called $c_p$ and frequency $\nu_p$~\citep{Simonovic:2017mhp}. We refer the reader to Appendix~\ref{app:fftlog} for more details on the FFTLog method, and to Appendix~\ref{app:Kernel} for the expressions of the coefficients $f_{nm}$.
This decomposition in powers of $k_1$ allows us to perform its integration analytically thanks to the following integral solutions \citep{DiDio:2015bua, DiDio:2018unb, Assassi:2017lea}
\begin{align}\label{eq:math}
     \int dk j_{\ell} (k\chi) j_{\ell} (kr) &= \frac{\pi}{2r^2}A_{\ell}(\chi, r)\,, \nonumber\\
     \int dk k^{2n} j_{\ell} (k\chi) j_{\ell} (kr) &= \frac{\pi}{2r^2} \left[-\frac{\partial^2}{\partial \chi^2} -\frac{2}{\chi} \frac{\partial }{\partial  \chi} + \frac{\ell(\ell+1)}{\chi^2}\right]^{n-1}\delta_{\rm D}(\chi-r)\,,\qquad n\geq 1 \nonumber\\
    \int dk k^{\nu - 1} j_{\ell} (k\chi) j_{\ell} (kr) &= \frac{\pi}{2r^2} I_{\ell}(\nu, r, \chi) \,, 
\end{align}
where $n$ is an integer and $\nu$ can be any complex number, and where we have defined the two functions
\begin{align}
     A_{\ell}(\chi, r_1) &= \frac{\chi}{1+2\ell} \left[ \left(\frac{r_1}{\chi}\right)^{\ell+2} \Theta(\chi-r_1)+ \left(\frac{r_1}{\chi}\right)^{-\ell+1} \Theta(r_1-\chi)\right]
     \,,\nonumber \\
     I_{\ell}(\nu, r_1, \chi) &=  \frac{r_1^2}{\chi^{\nu}} \frac{ 2^{\nu-2}\Gamma(\ell+\frac{\nu}{2})}{\Gamma(\frac{3-\nu}{2}) \Gamma(\ell+\frac{3}{2})} \left(\frac{r_1}{\chi}\right)^{\ell} {}_2F_1\left( \frac{\nu-1}{2}, \ell+\frac{\nu}{2}, \ell+\frac{3}{2}, \left(\frac{r_1}{\chi}\right)^2\right)\Theta(\chi-r_1) \nonumber\\
     &+  \frac{r_1^2}{\chi^{\nu}} \frac{ 2^{\nu-2}\Gamma(\ell+\frac{\nu}{2})}{\Gamma(\frac{3-\nu}{2}) \Gamma(\ell+\frac{3}{2})} \left(\frac{r_1}{\chi}\right)^{-\ell+1} {}_2F_1\left(\frac{\nu-1}{2}, \ell+\frac{\nu}{2}, \ell+\frac{3}{2}, \left(\frac{\chi}{r_1}\right)^{2}\right)\Theta(r_1-\chi)\,.
\end{align}
Following a similar computation as \cite{DiDio:2015bua, DiDio:2018unb} generalised to also involve relativistic terms, as well as defining the operator 
\begin{align}
    \mathcal{D}_\ell &= - \frac{\partial^2}{\partial \chi^2} +\frac{2}{\chi} \frac{\partial}{\partial \chi} + \frac{\ell \left( \ell+1 \right) -2 }{\chi^2}\,,
\end{align}
we find
\begin{align}\label{eq:density_term}
     b^{\delta_2}_{\ell_1 \ell_2 \ell_3} 
     &=
     2 \sum_{mn} \int d\chi C_{\ell_2}^{(n, 0)}(\chi)C_{\ell_3}^{(m, 0)}(\chi)\nonumber\\
     &
     \times \Bigg(  \chi^2\int \frac{dr_1}{r_1^{2}} D^2_{r_1}\tilde W_{r_1}   \Big[  f_{nm}^{(-4)}(r_1) I_{\ell_1}(-1, r_1,\chi)+f_{nm}^{(-2)}(r_1) A_{\ell_1}(r_1,\chi)  \nonumber\\
     &\qquad \left.+ \sum_p c_p \left(f_{nm}^{(-4, R)}(r_1) I_{\ell_1}(\nu_p-2, r_1,\chi) + f_{nm}^{(-2, R)}(r_1) I_{\ell_1}(\nu_p, r_1,\chi)\right) \Big]\right.\nonumber\\
     &
     \qquad+ f_{nm}^{(0)}(\chi)D^2_\chi\tilde W_\chi +  \mathcal D_{\ell_1}\left[f_{nm}^{(2)}(\chi)D^2_\chi\tilde W_\chi\right] + \mathcal D_{\ell_1}^2\left[f_{nm}^{(4)}(\chi)D^2_\chi\tilde W_\chi\right]  \Bigg)+ 2 \times \circlearrowleft 
          \,,
\end{align}
where the symbol $\circlearrowleft$ denotes the additional permutations of the $3$ multipoles, and the frequencies $\nu_p$ read $\nu_p = b+i\eta_p + 1$, with $b$ and $\eta_p$ the bias and the fundamental frequency of the FFTLog decomposition, which is described in more detail in Appendix~\ref{app:fftlog}. 
All these functions where defined by~\cite{Assassi:2017lea} and arise after having performed the analytical integration over $k_1$. The remaining integral over $r_1$ in Eq.\,\eqref{eq:density_term} is due to GR effects (second line) and radiation effects (third line). In the Newtonian limit, only the last three terms in the last line survive.

A similar computation for the second-order term $\partial_r^2 v_2$ leads to 
\begin{multline}\label{eq:G2}
     b^{\partial_r^2 v_2}_{\ell_1 \ell_2 \ell_3}  =
     - 2 \sum_{mn} \int d\chi\, C_{\ell_2}^{(n, 0)}(\chi)C_{\ell_3}^{(m, 0)}(\chi) 
     \left(\chi^2\int \frac{dr_1}{r_1^{2}} \left[\frac{d^2}{dr_1^{2}} \left[D^2_{r_1}\tilde W_{r_1} f_{nm}^{(-2)}(r_1)\right] I_{\ell_1}(-1, r_1,\chi) \right.\right.\\ 
      + \sum_p c_p \left(\frac{d^2}{dr_1^{2}}\left[D^2_{r_1}\tilde W_{r_1} f_{nm}^{(-2, R)}(r_1)\right] I_{\ell_1}(\nu_p-2, r_1,\chi) + \frac{d^2}{dr_1^{2}}\left[D^2_{r_1}\tilde W_{r_1} f_{nm}^{(-4, R)}(r_1)\right] I_{\ell_1}(\nu_p-4, r_1,\chi)\right) \\
        \left.+ \frac{d^2}{dr_1^{2}}\left[D^2_{r_1}\tilde W_{r_1} f_{nm}^{(0)}(r_1)\right] A_{\ell_1}(r_1,\chi) \right]
     \left.+ \frac{d^2}{d\chi^2} \left[f_{nm}^{(2)}(\chi)D^2_{\chi}\tilde W_\chi\right] +  \mathcal D_{\ell_1}\left[ \frac{d^2}{d\chi^2} \left[f_{nm}^{(4)}(\chi)D^2_\chi\tilde W_\chi\right]\right]\right)+ 2 \times \circlearrowleft\,. 
\end{multline}
The additional factor of $k_1^{-2}$ in the velocity (see Eq.\,\ref{eq:def_v2}) shifts the powers of $k_1$. Consequently, one of the Newtonian terms (third line) now involves an integral over $r_1$ that does not simplify further. Moreover, taking into account that $f_{nm}^{(-4)}=0$ (see Eqs.\,\ref{eqs:fnl_coef} in Appendix~\ref{app:Kernel}), this leaves only one pure GR contribution (first line) and a radiation contribution, analogous to Eq.\,\eqref{eq:density_term}, in the second line.

The last purely second-order term $\partial_r v_2$, see first term of Eq.\,\eqref{eq:projection}, is a projection effect. Its computation is similar to $\partial_r^2 v_2$, and leads to 
\begin{align}\label{eq:projection1}
     &b^{\partial_r v_2}_{\ell_1 \ell_2 \ell_3} = 
     2 \sum_{mn} \int d\chi C_{\ell_2}^{(n, 0)}(\chi)C_{\ell_3}^{(m, 0)}(\chi) \Bigg\{ \chi^2\int \frac{dr_1}{r_1^{2}} \Bigg[ \frac{d}{dr_1} \left[D^2_{r_1}\tilde W_{r_1} f_{nm}^{(-2)}(r_1) \mathcal R_{r_1} \mathcal H_{r_1} \right] I_{\ell_1}(-1, r_1,\chi) \nonumber \\
      &\qquad+ \sum_p c_p \left(\frac{d}{dr_1}\left[D^2_{r_1}\tilde W_{r_1} f_{nm}^{(-2, R) }(r_1) \mathcal R_{r_1} \mathcal H_{r_1}\right] I_{\ell_1}(\nu_p-2, r_1,\chi)\right. \nonumber\\
      &\qquad\qquad\qquad\left.+\frac{d}{dr_1}\left[D^2_{r_1}\tilde W_{r_1} f_{nm}^{(-4, R) }(r_1) \mathcal R_{r_1} \mathcal H_{r_1}\right] I_{\ell_1}(\nu_p-4, r_1,\chi) \right) \nonumber\\
     &\qquad+ \frac{d}{dr_1}\left[D^2_{r_1}\tilde W_{r_1} f_{nm}^{(0)} (r_1)\mathcal R_{r_1} \mathcal H_{r_1} \right] A_{\ell_1}(r_1,\chi) \Bigg] \nonumber\\
     &\qquad +  \frac{d}{d\chi} \left[ f_{nm}^{(2)}(\chi)D^2_{\chi}\tilde W_{\chi}\mathcal R_{\chi} \mathcal H_{\chi} \right] +    \mathcal  D_{\ell_1}\left[ \frac{d}{d\chi} \left[f_{nm}^{(4)}(\chi)D^2_{\chi}\tilde W_{\chi} \mathcal R_{\chi} \mathcal H_{\chi} \right]\right]\Bigg\}+ 2 \times \circlearrowleft \,.
\end{align}
It has a very similar structure as Eq.\,\eqref{eq:G2} with one pure GR term on the first line and one radiation term on the third and fourth lines. 

\subsection{Quadratic terms}
The quadratic terms are simpler to compute, since their kernels are trivial; they have no residual momentum dependence. Hence, the analytic integration over $k_1$ gives a Dirac delta from the second equation of \eqref{eq:math} where $n=1$. Apart from $\partial_\alpha \partial^\alpha \partial_r v$, all quadratic terms read as an integral over $r_1$ of some function of $r_1$ times two generalised power spectra. Thanks to Eqs.\,\eqref{eqs:generalized_ps_def}, we can write all quadratic terms in a very compact way
\begin{align}\label{eq:quadratic}
   b^{XY}_{\ell_1 \ell_2 \ell_3} 
    &=  \int dr_1   \tilde W_{r_1} C^{X}_{\ell_2}(r_1) C^{Y}_{\ell_3}(r_1) + 5\times \circlearrowleft\,,
\end{align}
where $X$ and $Y$ can be any superscript of the left-hand side of Eqs.\,\eqref{eqs:generalized_ps_def}. 
Quadratic Newtonian terms were previously computed without employing the Limber approximation by~\cite{DiDio:2015bua}. However, since FFTLog techniques were not used there, the numerical evaluation was significantly slower than in our work.

The last term involves angular derivatives, which should be treated in spherical harmonics space,
\begin{eqnarray}\label{eq:davd1v}
    b^{\partial_\alpha v \partial^\alpha \partial_r v}_{\ell_1 \ell_2 \ell_3} &=& 2 \sqrt{\ell_2(\ell_2+1)}\sqrt{\ell_3(\ell_3+1)} \mathcal A_{\ell_1\ell_2\ell_3}
     \int dr_1  \tilde W_{r_1} f_{r_1}C_{\ell_2}^{(-2, 0)} C_{\ell_3}^{\partial_r v}
     + 5\times \circlearrowleft \,,
\end{eqnarray}
where the geometric factor $\mathcal A_{\ell_1\ell_2\ell_3}$ is defined as 
\citep{DiDio:2015bua}
\be\label{eq:Alll}
\mathcal A_{\ell_1\ell_2\ell_3} = \frac{1}{2}
\left[
\begin{pmatrix}
\ell_1 & \ell_2 & \ell_3 \\
0 & 1 & -1
\end{pmatrix}+
\begin{pmatrix}
\ell_1 & \ell_2 & \ell_3 \\
0 & -1 & 1
\end{pmatrix}
\right] \Bigg/
\begin{pmatrix}
\ell_1 & \ell_2 & \ell_3 \\
0 & 0 & 0
\end{pmatrix}
\,.
\ee
Given the above simplifications, the numerical evaluation of each bispectrum term is feasible. We have numerically implemented Eqs.\,\eqref{eq:density_term}--\eqref{eq:davd1v} in the Python code \href{https://github.com/TomaMTD/ang_bispec}{\texttt{ang\_bispec}} used to produce the results of this paper. The code structure is briefly described in Appendix~\ref{app:code}. In the next section, we present the first results, focusing on the comparison of the different terms.

\section{Analytical results}\label{sec:Theoretical results}

\begin{figure}
    \centering
    \includegraphics[scale=0.47]{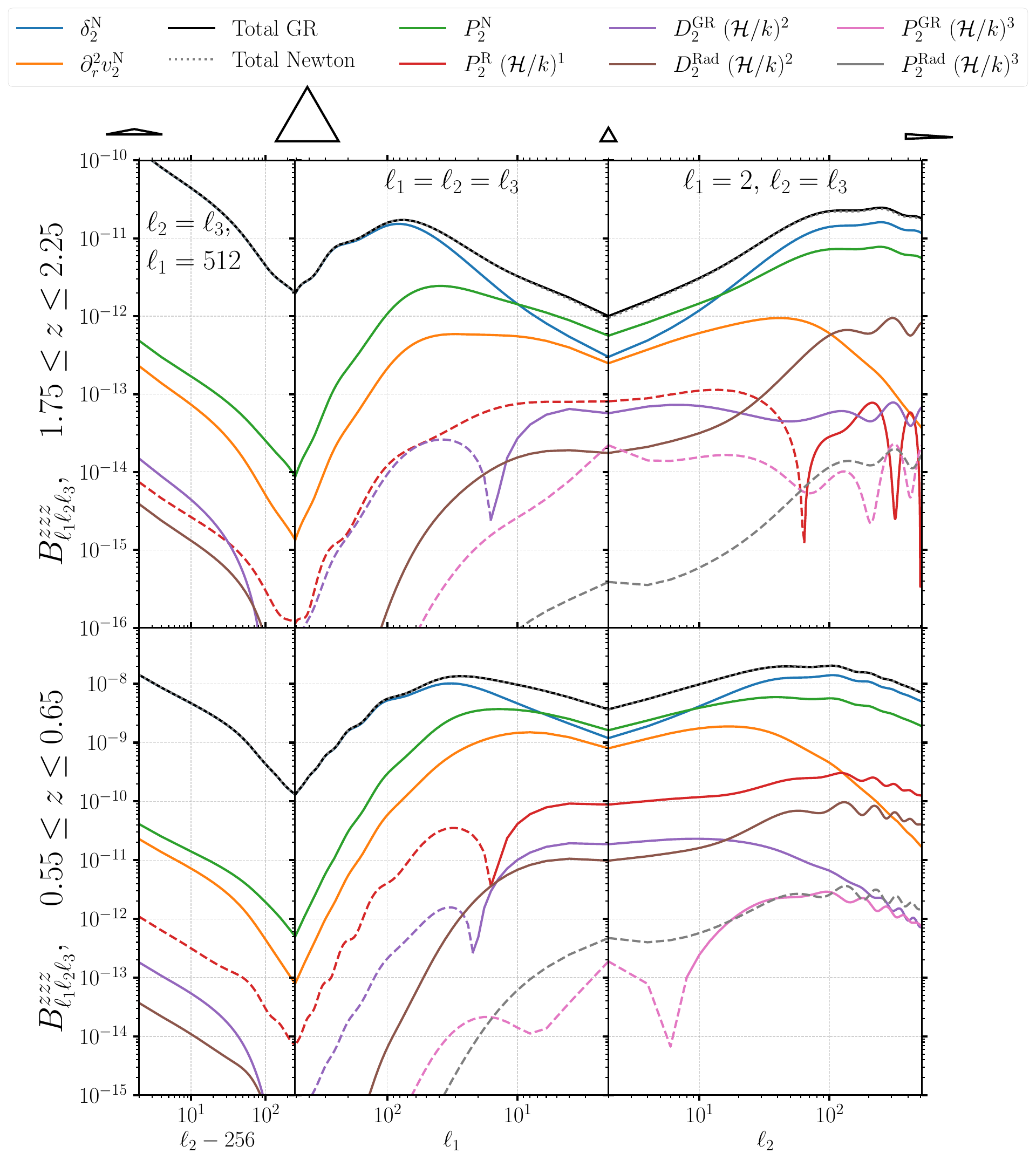}
    \caption{We show the angular bispectrum for two redshift bins: $1.75 \leq z \leq 2.25$ on the first row and $0.55 \leq z \leq 0.65$ on the second row. The first column, we have fixed $\ell_1=514$ and vary $\ell_2=\ell_3$ on the $x$-axis. The middle column shows equilateral configurations. Note the reversed $x$-axis which ensure the continuity of the curves between the columns. In the right column, we have fixed $\ell_1=2$ and increase $\ell_2=\ell_3$. The grey dotted line represents the sum of all the terms, including all relativistic terms, while the black line only contains the Newtonian terms indicated in the legend with a ``N" subscript. The relativistic projection effects are shown in red while the pure dynamical GR (radiation) effects are shown in brown (violet). In grey and pink, we show the coupling between projection relativistic effects with GR and radiation, respectively. Dashed lines indicates negative values.}
    \label{fig:term_comparison}
\end{figure}
\begin{figure}
    \centering
    \includegraphics[scale=0.5]{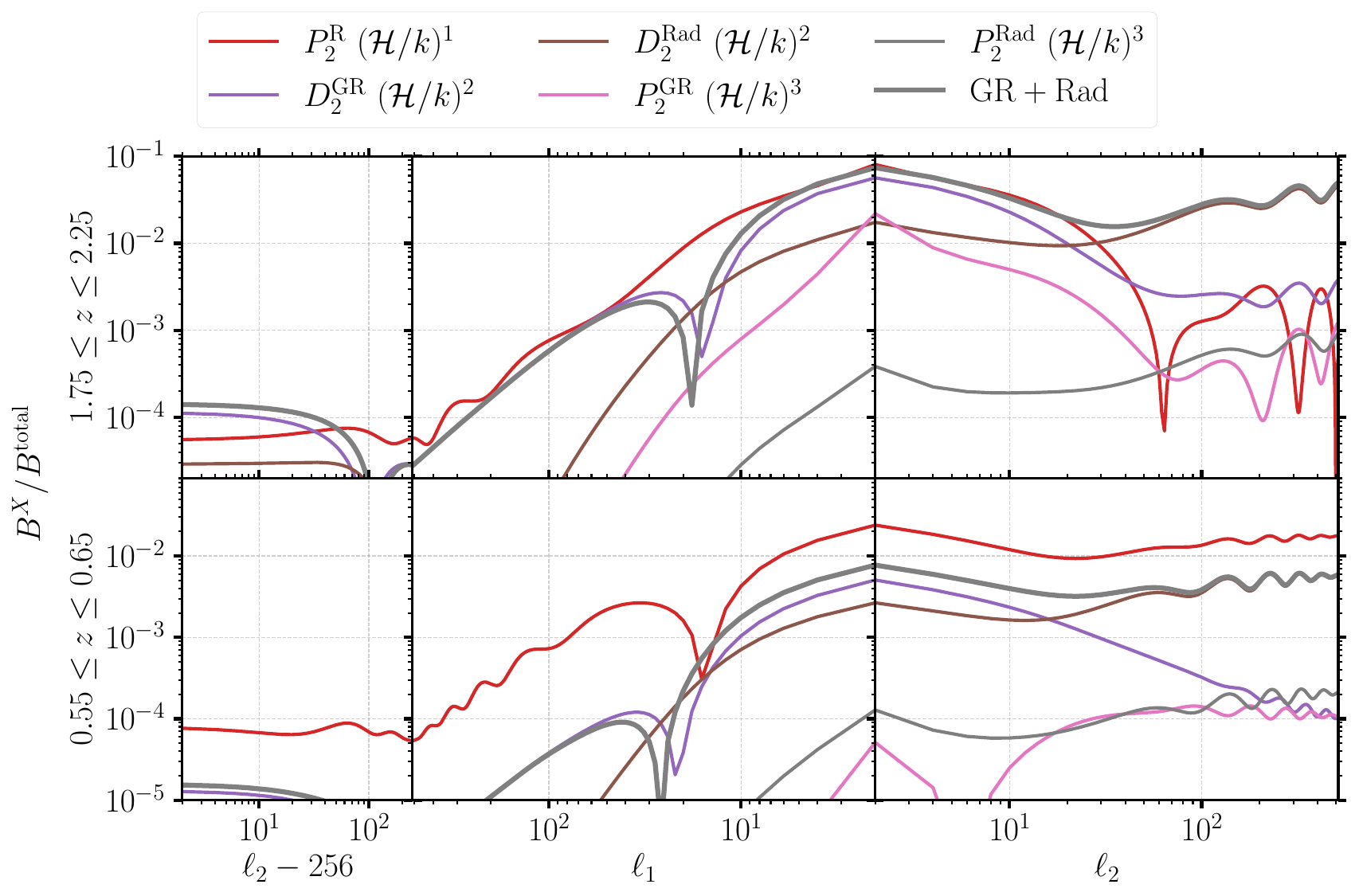}
    \caption{We plot the ratio between the relativistic contributions and the total bispectrum shown in Fig.\,\ref{fig:term_comparison}. The structure of the panel is the same as Fig.\,\ref{fig:term_comparison}. We can see that the relativistic contributions reach almost $10 \%$ of the total amplitude at $1.75 \leq z \leq 2.25 $ for large-scale equilateral configurations and the squeezed limit. At $0.55 \leq z \leq 0.65 $, the relative amplitude reaches at most  $2\%$.}
    \label{fig:ratio}
\end{figure}

As noted earlier, our bispectrum is derived through numerical integration of the theoretical expressions up to second order in perturbation theory.
We begin by analysing the different contributions to the theoretical bispectrum. 
For convenience, we regroup the 14 bispectrum contributions in Eq.\,\eqref{eq:fullbl} into $8$ terms, each of them having distinct physical origin and/or have comparable order of magnitude: 
\begin{align}\label{eq:b_split}
    B^{\hat z_1 \hat z_2 \hat z_3}_{\ell_1\ell_2\ell_3} &= B_{\ell_1\ell_2\ell_3}^{\delta_2^{\mathrm N}} + B_{\ell_1\ell_2\ell_3}^{\partial_r^2 v_2^{\mathrm N}} + B_{\ell_1\ell_2\ell_3}^{P_2^{\mathrm N}} + B_{\ell_1\ell_2\ell_3}^{P_2^{\mathrm R}}+ B_{\ell_1\ell_2\ell_3}^{D_{2}^{\mathrm {GR}}}+ B_{\ell_1\ell_2\ell_3}^{D_{2}^{\mathrm {Rad}}}+ B_{\ell_1\ell_2\ell_3}^{P_{2}^{\mathrm {GR}}}+ B_{\ell_1\ell_2\ell_3}^{P_{2}^{\mathrm {Rad}}}\,,
\end{align}
where, as before we suppress redshift dependencies for notational ease. To understand better the link between the terms of Eq.\,\eqref{eq:b_split} and our starting expressions for $\Delta$, we define the spherical harmonics projection operator (from Eqs.~\ref{eq:angleaverage} and~\ref{eq:alm_def}), 
\begin{equation}
    \langle \cdots \rangle_{\cal S}
    :=  N_{\ell_1 \ell_2 \ell_3} \! \sum_{m_1m_2m_2} 
    \begin{pmatrix}
        \ell_1 & \ell_2 & \ell_3\\
        m_1 & m_2 & m_3
    \end{pmatrix}
    \int \! d\Omega_{\boldsymbol{n}_1} d\Omega_{\boldsymbol{n}_2} d\Omega_{\boldsymbol{n}_3} Y^*_{\ell_1m_1}(\boldsymbol{n}_1) Y^*_{\ell_2m_2}(\boldsymbol{n}_2) Y^*_{\ell_3m_3}(\boldsymbol{n}_3) \langle \cdots \rangle ,
\end{equation}
such that we can directly link the Fourier-space correlator of the fields of section~\ref{sec:Number Counts} with the angular bispectrum. Hence, we have 
\begin{equation}
   B_{\ell_1\ell_2\ell_3}^{z_1z_2z_3} = 
   \langle \Delta(z_1, \boldsymbol{n}_1) \Delta(z_2, \boldsymbol{n}_2) \Delta(z_3, \boldsymbol{n}_3) \rangle_{\cal S} \,.
\end{equation}
With this prelude, the ``Newtonian'' terms of Eq.\,\eqref{eq:b_split} read at ${\cal O}(\mathcal H^0/k^0$):
\begin{align}
    B^{\delta_2^{\rm N}}_{\ell_1\ell_2\ell_3} = 
    \langle \delta_2^{\rm N} \Delta_1^{\rm N} \Delta_1^{\rm N} \rangle_{\cal S} \,, \quad
    B^{\partial_r^2 v_2^{\rm N}}_{\ell_1\ell_2\ell_3}  = 
    \langle \partial_r^2 v_2^{\rm N} \Delta_1^{\rm N} \Delta_1^{\rm N} \rangle_{\cal S}\,,
    \quad
    B^{P_2^{\rm N}}_{\ell_1\ell_2\ell_3} = 
    \langle P_2^{\rm N} \Delta_1^{\rm N} \Delta_1^{\rm N} \rangle_{\cal S} \,,
\end{align}
and the relativistic projection effects at ${\cal O}({\cal H}/k)$ are
\be
 B^{P_2^{\rm R}}_{\ell_1\ell_2\ell_3} = 
 \langle P_2^{\rm R} \Delta_1^{\rm N} \Delta_1^{\rm N} \rangle_{\cal S} + \langle \Delta_2^{\rm N} P_1^{\rm R} \Delta_1^{\rm N} \rangle_{\cal S} \,. 
\ee
In contrast, the dynamical GR and radiation effects at ${\cal O}({\cal H}^2/k^2)$ are respectively
\begin{align}
    B^{D_2^{\rm GR}}_{\ell_1\ell_2\ell_3} = 
    \langle D_2^{\rm GR} \Delta_1^{\rm N} \Delta_1^{\rm N} \rangle_{\cal S} + \langle \Delta_2^{\rm N} D_1^{\rm GR} \Delta_1^{\rm N} \rangle_{\cal S}
    \,, \qquad
    B^{D_2^{\rm Rad}}_{\ell_1\ell_2\ell_3} = 
    \langle D_2^{\rm Rad} \Delta_1^{\rm N} \Delta_1^{\rm N} \rangle_{\cal S} \,,
\end{align}
while couplings between relativistic projection and  radiation contributions at ${\cal O}(\mathcal H^3/k^3)$ read
\begin{align}
    B^{P_2^{\rm GR}}_{\ell_1\ell_2\ell_3} &= 
    \langle P_2^{\rm GR} \Delta_1^{\rm N} \Delta_1^{\rm N} \rangle_{\cal S} + \langle P_2^{\rm R} D_1^{\rm GR} \Delta_1^{\rm N} \rangle_{\cal S} + \langle D_2^{\rm GR} P_1^{\rm R} \Delta_1^{\rm N} \rangle_{\cal S}
    \,,\\
    B^{P_2^{\rm Rad}}_{\ell_1\ell_2\ell_3} &= 
    \langle P_2^{\rm Rad} \Delta_1^{\rm N} \Delta_1^{\rm N} \rangle_{\cal S} + \langle D_2^{\rm Rad} P_1^{\rm R} \Delta_1^{\rm N} \rangle_{\cal S} \,,
\end{align}
and we remind the reader that all appearing first-order (second-order) terms are defined in Sec.\,\ref{sec:FO} (Sec.\,\ref{sec:SO}).

In Fig.\,\ref{fig:term_comparison}, we plot each term of Eq.\,\eqref{eq:b_split} for two redshift bins: $1.75 \leq z \leq 2.25$ in the first row and $0.55 \leq z \leq 0.65$ in the second row. The smoothness parameter $\sigma$ defined in Eq.\,\eqref{eq:window} is chosen as a compromise: it must be small enough to satisfy the condition $\sigma \ll r_{\rm max} - r_{\rm min}$, ensuring a close match to a top-hat window, but large enough for the analytical derivatives to be numerically stable and well sampled. We find a good compromise with the criterion $\sigma / [r_{\rm max} - r_{\rm min}] \simeq 0.05$. 
For $1.75 \leq z \leq 2.25$ we set $\sigma=25$~Mpc$/h$ while for $0.55 \leq z \leq 0.65$ we set $ \sigma=10$~Mpc$/h$.
We indicate positive (negative) values with solid (dashed) lines. In black (dotted grey) we show the total GR (Newtonian) bispectrum. In the left column, we have fixed one multipole to a large value $\ell_1=512$ and vary $\ell_2=\ell_3$ so that the first point corresponds to a folded configuration with $2\ell_2=2\ell_3=\ell_1$ while the last point is an equilateral configuration that corresponds to the starting configuration of the second column. The middle column shows the equilateral configurations from $\ell_1=514$ to $\ell_1=2$ (note the flipped horizontal axis, with $\ell_1$ decreasing towards the right). Hence, the last point of the middle column corresponds to a large-scale triangle connecting with the first point of the third column where we fix $\ell_1=2$ and vary $\ell_2=\ell_3$. Therefore, the right part of the last column corresponds to squeezed triangles $\ell_2=\ell_3 \gg \ell_1$. 

In Fig.\,\ref{fig:ratio}, we show ratios between the relativistic bispectra and the total bispectrum; the structure of the panels and the colours correspond to those of Fig.\,\ref{fig:term_comparison}. For the bin $\hat z: \ 1.75 \leq z \leq 2.25$, the various relativistic effects comprise almost 10\% (5\%) of the total bispectrum signal  on large scales for equilateral (squeezed) configurations; these amplitude results are slightly larger as obtained by the simulation measurements of \cite{Montandon:2022ulz} (see their Fig.~13). In contrast, for the bin $\hat z: 0.55 \leq z \leq 0.65$, the ratios peak at about 1\% at large scales, which agrees fairly well with \cite{Montandon:2022ulz}.

It is worth noting that the width of the redshift window functions corresponds to a characteristic scale below which the power is being suppressed. This suppression is caused by the integration on the line of sight, which mixes different scales. Indeed, a given multipole corresponds to different scales at different redshifts. For scales much larger than the redshift bin width, the different can be neglected. For scales much smaller, it results in a loss of correlation, and hence of power. The width of the redshift bins has been adjusted to correspond to the same multipoles, roughly corresponding to $\ell \sim 45$.

Let us now discuss the various contributions.

\begin{figure}
    \centering
    \includegraphics[scale=0.5]{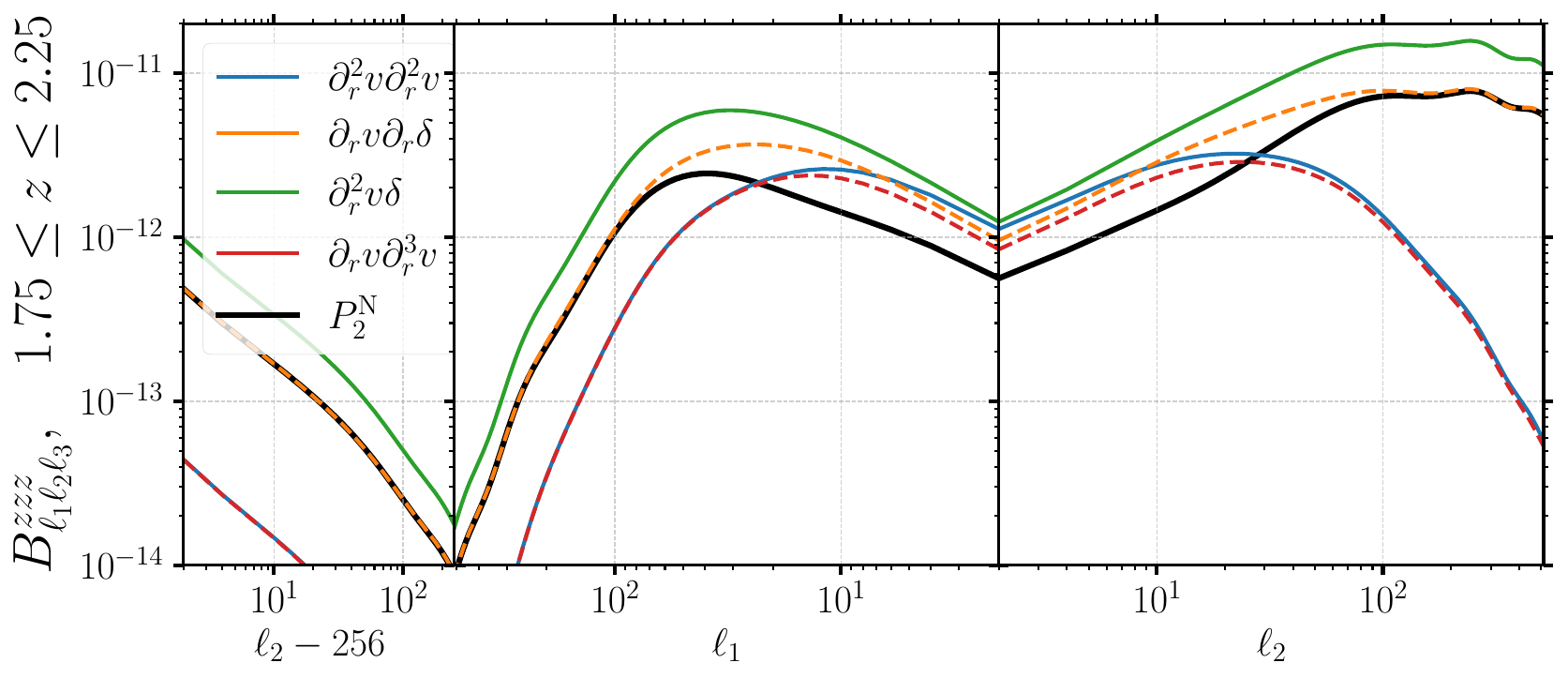}
    \caption{We show each contribution to the Newtonian quadratic term $P^{\rm N}_2$ according to \eqref{eq:QN} for the redshift bin $1.75 \leq z \leq 2.25 $. The axis are the same as Fig.\,\ref{fig:term_comparison}.}
    \label{fig:quadratic}
\end{figure}

\vbox{
\subsection{Newtonian terms} 

The Newtonian terms $\delta_2^{\mathrm N}$, $\partial_r^2 v_2^{\mathrm N}$, and $P_2^{\mathrm N}$ are shown, respectively, in blue, orange, and green. As expected, they are the dominant terms, with the density term dominating at small scales except close to the large-scale equilateral configuration where the quadratic Newtonian term dominates. The RSD term in orange is the smallest Newtonian contribution. In the squeezed limit, it decays and becomes smaller even in relation to relativistic terms. In Fig.\,\ref{fig:quadratic}, we show the four terms included in $P_{2}^{\rm N}$, shown in black. We find an interesting cancellation between $\partial_r^2 v \partial_r^2 v$ and $\partial_r v \partial_r^3 v$ for all configurations considered except for the large-scale equilateral configuration. For the two density terms and for the same configurations, we see again a partial cancellation, but now with a factor $2$ offset, that is, $B^{\partial_r^2 v \delta} \simeq -2 B^{\partial_r v \partial_r \delta}$. 
Hence, for most configurations that involve at least one small scale, we have $B^{P_2^{\rm N}} \simeq -B^{\partial_r v \partial_r \delta}$. We refer the reader to the Appendix \ref{app:cancellation} for a discussion about these cancellations. 
}

\subsection{Projection effects} 
In Figs.~\ref{fig:term_comparison} and \ref{fig:ratio}, the relativistic projection effects at order $\mathcal H/k$, $P_2^{\rm R}$, are shown in red; see Eqs.\,\eqref{eq:projection}. At high redshift $z=2$, it has a negative contribution for all configurations except in the squeezed limit. The amplitude of its contribution is similar to that of GR effects. At $z=2$, the cosmological horizon corresponds to a multipole $\ell \simeq \HH r \simeq 1.2$. Hence, for the largest scales and in the squeezed limit, we are close to the regime $\mathcal H/k \sim 1$, which is why all relativistic terms are of the same order. However, for the lowest redshift, the modes probed are deeper inside the horizon such that the relativistic hierarchy in powers of $\mathcal H/k$ holds, which explains why relativistic projection effects dominate. The term $B^{P^{\rm R}_2}$ is the result of many contributions that are shown in Fig.\,\ref{fig:quadraticPR} and discussed in Appendix~\ref{app:PR}.

\begin{figure}
    \centering
    \includegraphics[scale=0.5]{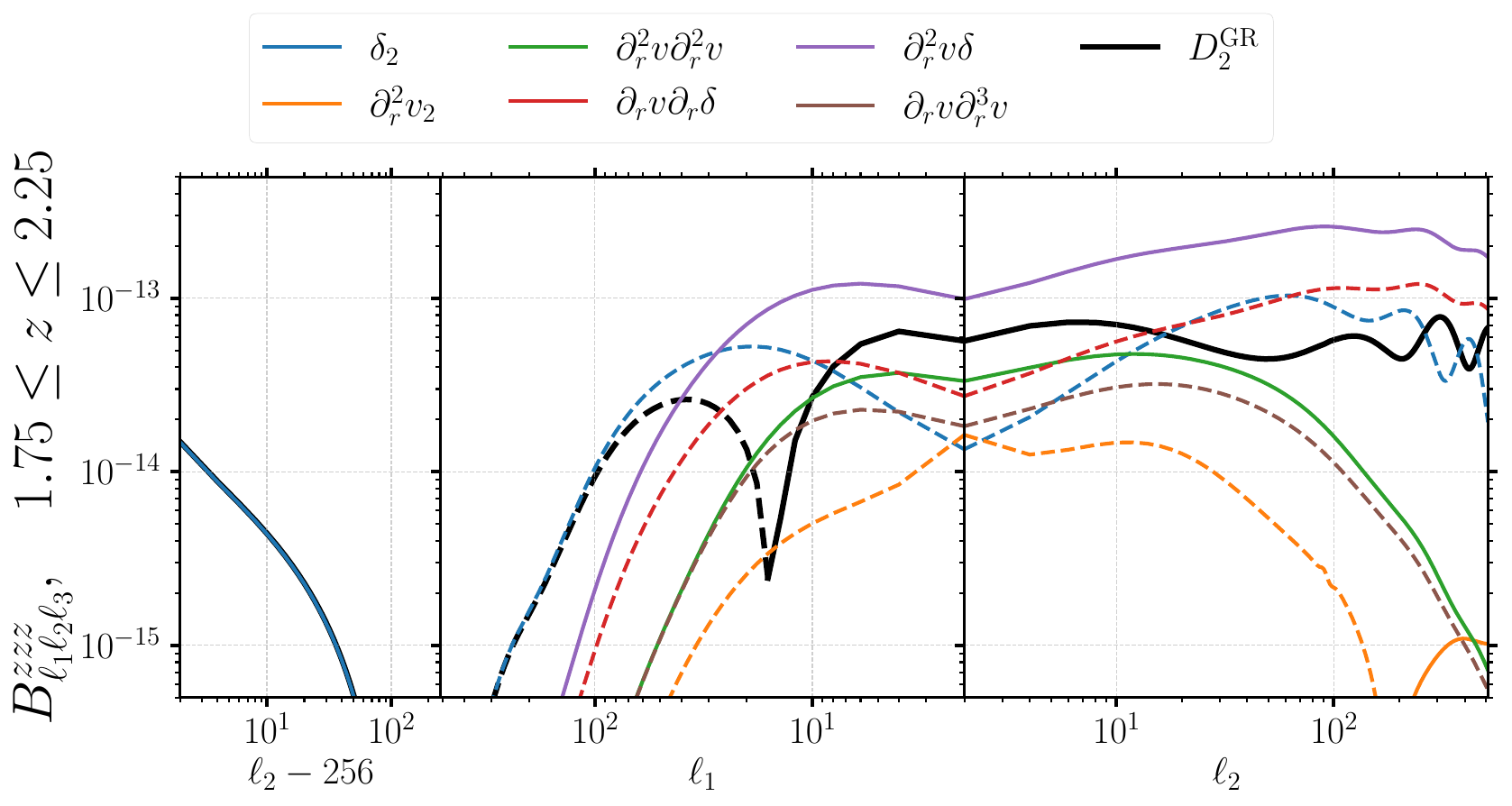}
    \caption{We show each contribution to the general relativistic term $D^{\rm GR}_2$ according to the first equation of \eqref{eq:gr} for the redshift bin $1.75 \leq z \leq 2.25$. We consider here only the dominant terms, hence neglecting the coupling of projection effects with GR terms. 
    The axes are the same as Fig.\,\ref{fig:quadratic}.}
    \label{fig:quadraticGR}
\end{figure}

\subsection{General relativistic effects} 

The pure GR effects are represented in violet in Figs.~\ref{fig:term_comparison} and \ref{fig:ratio}. At $z=2$, its contribution is very similar to projection effects, except that it has the opposite sign for large-scale equilateral triangles and folded configurations. In the squeezed limit, it does not decay and oscillates out of phase with respect to the projection effects. At $z=0.6$, it becomes, as expected, smaller than the projection effects and seems to decay in the squeezed limit, becoming of the same order as the $(\mathcal H / k)^3$ terms. In Fig.\,\ref{fig:ratio}, we can see that in the large-scale equilateral limit and at $z=2$, the GR effects represent $5$-$6\%$ of the total amplitude, close to the relativistic projection effects which are about $7$-$8\%$. In the squeezed limit, they drop with the projection effects to $0.3$-$0.4\%$. 

In Fig.\,\ref{fig:quadraticGR}, we show all the contributions to $D^{\rm GR}_2$ according to Eq.\,\eqref{eq:gr}. We can see that the GR effects induced by the RSD term in orange are generally negligible. The cancellations involving quadratic terms, as discussed before, still hold meaning that for small-scale equilateral, folded, and squeezed configurations, the GR effects coming from the quadratic terms are roughly $\simeq - B^{\partial_r v \partial_r \delta}_{{\rm GR}}$. However, the density term dominates in the small-scale equilateral and folded configurations such that we can here neglect the quadratic terms. In the squeezed limit, GR effects can be approximated by $B^{\delta_2^{\rm GR}} - B^{\partial_r v \partial_r \delta}_{{\rm GR}}$.

\begin{figure}
    \centering
    \includegraphics[scale=0.5]{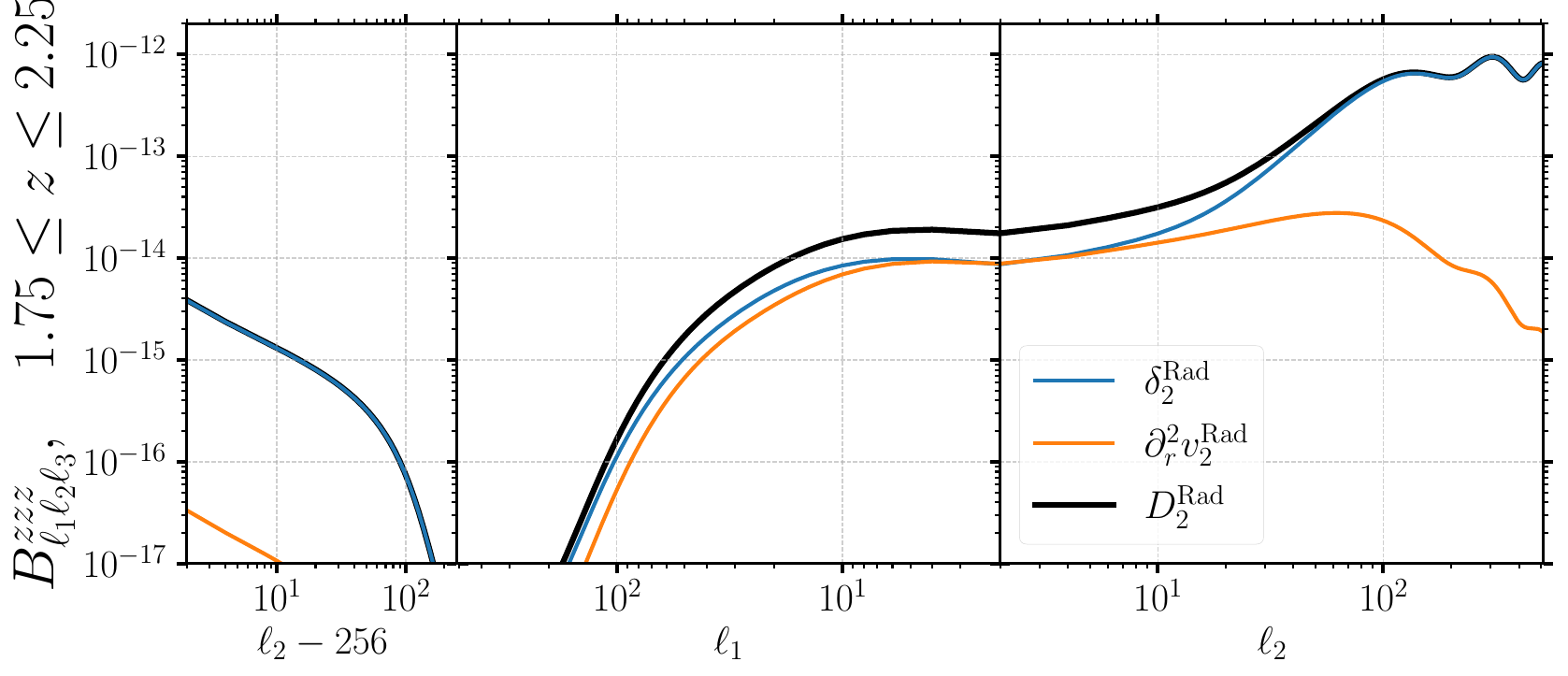}
    \caption{We show each contribution to the radiation term $D^{\rm Rad}_2$ according to \eqref{eq:gr} for the redshift bin $1.75 \leq z \leq 2.25$. We consider here only the dominant terms, hence neglecting the coupling of projection effects with radiation terms.}
    \label{fig:quadraticRad}
\end{figure}

\subsection{Radiation} 
The radiation term, shown in brown in Figs.~\ref{fig:term_comparison} and \ref{fig:ratio}, is subdominant for all configurations except in the squeezed limit where it becomes larger than the (other) GR effects. At $z=2$, it is the dominant relativistic contribution and even $10\times$ larger than the Newtonian RSD term and the other relativistic terms. For $z=0.6$, it remains always smaller than projection effects, but the relative amplitude is about $30\%$. The radiation approximation that we have used consists of a correction in the squeezed limit where most of radiative effects are expected to have a sizable effect~\cite[see][]{Tram:2016cpy}. In Fig.\,\ref{fig:quadraticRad}, we show the two contributions to $B^{D^{\rm Rad}}_2$. Hence we can say that in the squeezed limit and for folded configurations, the density terms dominate, whereas for equilateral configurations, both contributions are similar.   

\subsection{Terms $\propto (\mathcal H / k)^3$}
Finally, in Figs.~\ref{fig:term_comparison} and \ref{fig:ratio}, shown in in pink and grey, we show respectively the GR and radiation effects on the relativistic projection effects. The radiation effect roughly follows the main radiation part, but roughly two orders of magnitude smaller at $z=2$ and one order of magnitude at $z=0.6$. Compared to the relativistic projection effects, both the radiation and the GR part represent in a squeezed limit a $\sim 10\%$ effect at $z=2$ and $\sim 1\%$ at $z=0.6$. 

To conclude, we find that the Newtonian density is the dominant term for small-scale equilateral/folded configurations. At large scales and in the squeezed limit, the amplitude of Newtonian quadratic terms becomes of the same order as the density. We find an interesting cancellation within the Newtonian quadratic terms such that for squeezed configurations we can write 
\begin{equation}
 B_{\rm squeezed} \simeq  B^{\delta} - B^{\partial_r v \partial_r \delta}\,,
\end{equation}
which holds for both Newtonian and GR effects.

Moreover, we find that the effects of GR and radiation seem to be more important than expected. At redshift $z=2$, GR effects are comparable to relativistic projection effects, which should be an order $\mathcal H/ k$ smaller, while the radiation effect dominates by one order of magnitude. However, note that the horizon at $z=2$ is, roughly speaking, located around $\ell=2$. This means that for the largest scales studied at this redshift, the weak-field expansion parameter $\mathcal{H}/k$ becomes of order unity. For the smaller scales, it also strongly depends on the kernels' momentum dependence. Hence, in the squeezed limit at low redshift, it represents more than $\sim 30\%$ of the amplitude of the relativistic effects. In general, with Fig.\,\ref{fig:ratio}, we have confirmed the amplitude estimated with simulations by~\cite{Montandon:2022ulz}. Moreover, the projection effects start to appear at first order in the weak-field expansion $\HH/k$. This implies that, when computing the bispectrum of these terms, we encounter seven spatial derivatives of the gravitational potential. With an odd number of derivatives, there will always be at least one $k$-integral involving two spherical Bessel functions that oscillate out of phase, suppressing the bispectrum amplitude with respect to a simple derivative power counting.

We now compare our analytical results with the simulations that were performed by~\cite{Montandon:2022ulz}. In simulations, however, all terms are mixed together and, in general, we only have access to the total bispectrum. Hence, we will now only study comparisons between the sums of the terms and leave a more detailed analysis, potentially isolating individual terms in simulations, for future work.

\section{Comparison with simulation}\label{sec:Comparison with simulation}
To test our implementation of Eq.\,\eqref{eq:fullbl}, we compare our analytical calculation with the bispectrum measurements on the simulations that have been performed by~\cite{Montandon:2022ulz}. For this we have access to ten paired light cones~\cite[see][]{Angulo:2016hjd} for more details on the pairing method) that were simulated with relativistic dynamics and including radiation perturbations and ten paired light cones that were simulated with Newtonian dynamics.
In addition, we rerun the ray-tracing algorithm setting the metric potential $\phi$ (henceforth called `the potential') to zero. This way, the final bispectrum measurements on these light cones will be clean of any geometric effects due to the potential, in particular lensing terms and integrated Sachs-Wolfe effects which we do not compute in our analytical treatment. However, setting $\phi=0$, we also miss second-order gravitational redshifts and quadratic terms coupling the linear potential, such as the last three terms of Eq.\,\eqref{eqs:pot_terms}. Even though neglecting these contributions plays a subdominant role in the investigation of GR and radiation effects, we omit them in the theoretical computation for consistency with the simulation setup. 

Since relativistic effects play only a subdominant role compared to Newtonian contributions, the total bispectrum can be compared with both Newtonian and relativistic simulations. For this we choose to compare the sum of all terms in Eq.\,\eqref{eq:fullbl}, excluding the last three potential terms, with the bispectrum measured in the relativistic simulations. Regarding the relativistic effects, \cite{Montandon:2022ulz} estimated the relativistic contribution by subtracting the bispectrum measured in relativistic simulations from that measured in Newtonian simulations. As we will discuss in more detail, these measurements cannot be reduced to a ``pure'' dynamical relativistic effects.

For a proper comparison between the simulation measurements and the theory computed in this paper, we still miss two mandatory steps for the estimation of the bispectrum with the binned bispectrum estimator \citep{Bucher:2015ura}: binning and smoothing. First, the binning consists of averaging the bispectrum for close triangle configurations to increase the signal-to-noise ratio. This estimator computes the binned averaged angular bispectrum which can be expressed as a function of the reduced bispectrum as (see eq.\,\ref{eq:angleaverage})
\begin{equation}\label{eq:binned}
        B^{z_1z_2z_3}_{i_1i_2i_3} = \frac{1}{\Xi_{i_1i_2i_3}} \sum_{\ell_1 =\ell^{\rm min}_{i_1}}^{\ell_{i_1}^{\rm max}}\sum_{\ell_2 =\ell^{\rm min}_{i_2}}^{\ell_{i_2}^{\rm max}}\sum_{\ell_3 =\ell^{\rm min}_{i_3}}^{\ell_{i_3}^{\rm max}} B_{\ell_1\ell_2\ell_3}^{z_1z_2z_3} \,.
\end{equation}
Here, we have defined $i_1$, $i_2$, and $i_3$, which label the bins, $\Xi_{i_1i_2i_3}$ the number of triangles in a given bin and $\ell_{\rm min}$
and $\ell_{\rm max}$, which delimit the considered bin. Second, to increase the signal-to-noise ratio even more to have a good detection of the tiny relativistic effects, we convolve the bispectrum with a Gaussian 
\begin{equation}\label{eq:smooth}
    S\left[ B^{z_1 z_2 z_3}_{i_1 i_2 i_3}\right] =  
    (2\pi\sigma^2_{\rm bin})^{-3/2}\sum_{i_1',i_2',i_3'} \exp{\left[-\frac{1}{2} \frac{ (i_1 - i_1')^2+(i_2 - i_2')^2+(i_3-i_3')^2 }{\sigma_{\rm bin}^2} \right]} 
    B^{z_1z_2z_3}_{i_1'i_2'i_3'}\,,
\end{equation}
where $\sigma_{\rm bin}$ is the standard deviation of the Gaussian. In the following, we will compare the (non-)smoothed binned bispectrum estimated in the simulations with the (non-)smoothed and binned analytical computation. We should notice here that we neglect the numerical effects that affect the bispectrum measurements in the simulation. Indeed, as already found by~\cite{Montandon:2022ulz} for the power spectrum, at small scales due to the finite resolution, the power drops in the simulation compared to the theory. We expect the same effect in the bispectrum. Since we do not include this effect in the theory, the smoothing, which mixes different configurations, may create a general offset even at large scales between the theory and the measurements. The amplitude of this effect depends on the standard deviation $\sigma_{\rm bin}$. We have a good detection of the total bispectrum in the simulation, and hence the smoothing is not necessary. Hence, in this section we compare the binned total bispectrum without smoothing, and defer the discussion of smoothing effects to the appendix, since smoothing is essential for understanding the relativistic comparison, which cannot be carried out without it. 

\begin{figure}
    \centering
    \includegraphics[scale=0.5]{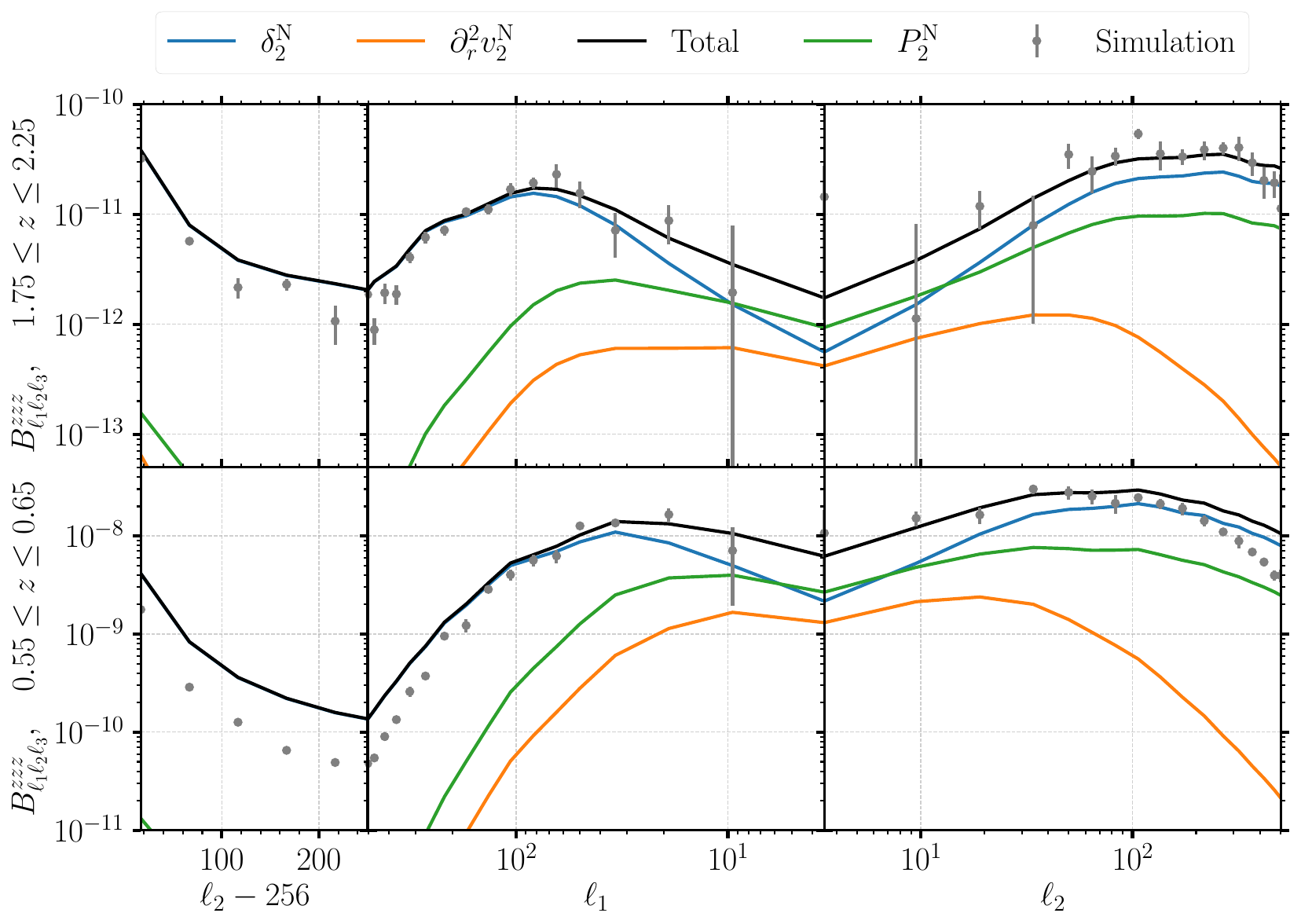}
    \caption{We show here the main contributions to the binned bispectrum and compare them to measurements from simulations (grey points with error bars). Panels are similar to Fig.\,\ref{fig:term_comparison}.}
    \label{fig:simVSth}
\end{figure}

\begin{figure}
    \centering
    \includegraphics[scale=0.5]{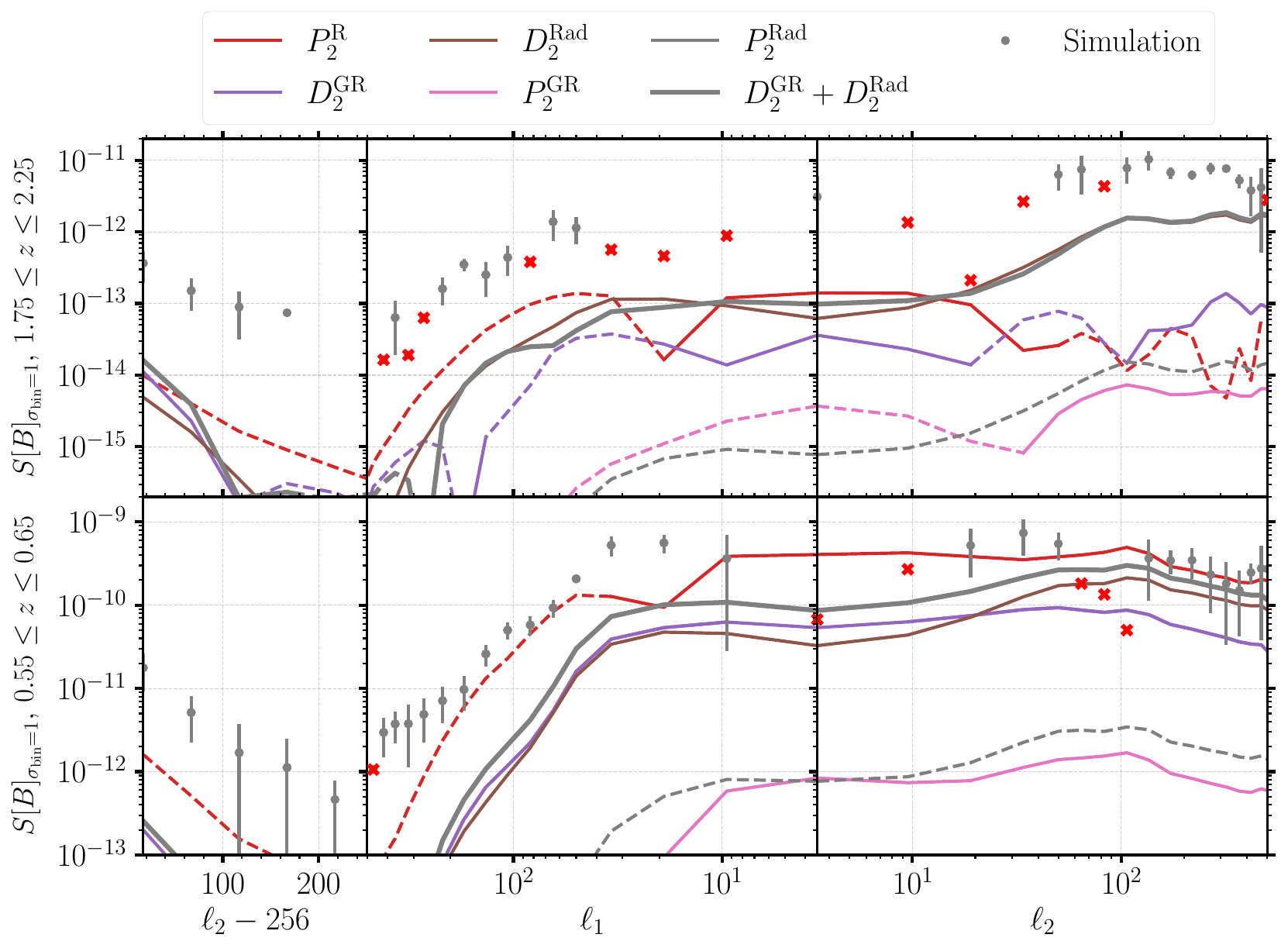}
    \caption{We show here the relativistic contributions to the binned smooth bispectrum and compare them to measurements from simulations (grey points with error bars). The red crosses indicate large error bars that cross zero which have been removed for readability. }
    \label{fig:simVSth_rela}
\end{figure}

\begin{figure}
    \centering
    \includegraphics[scale=0.48]{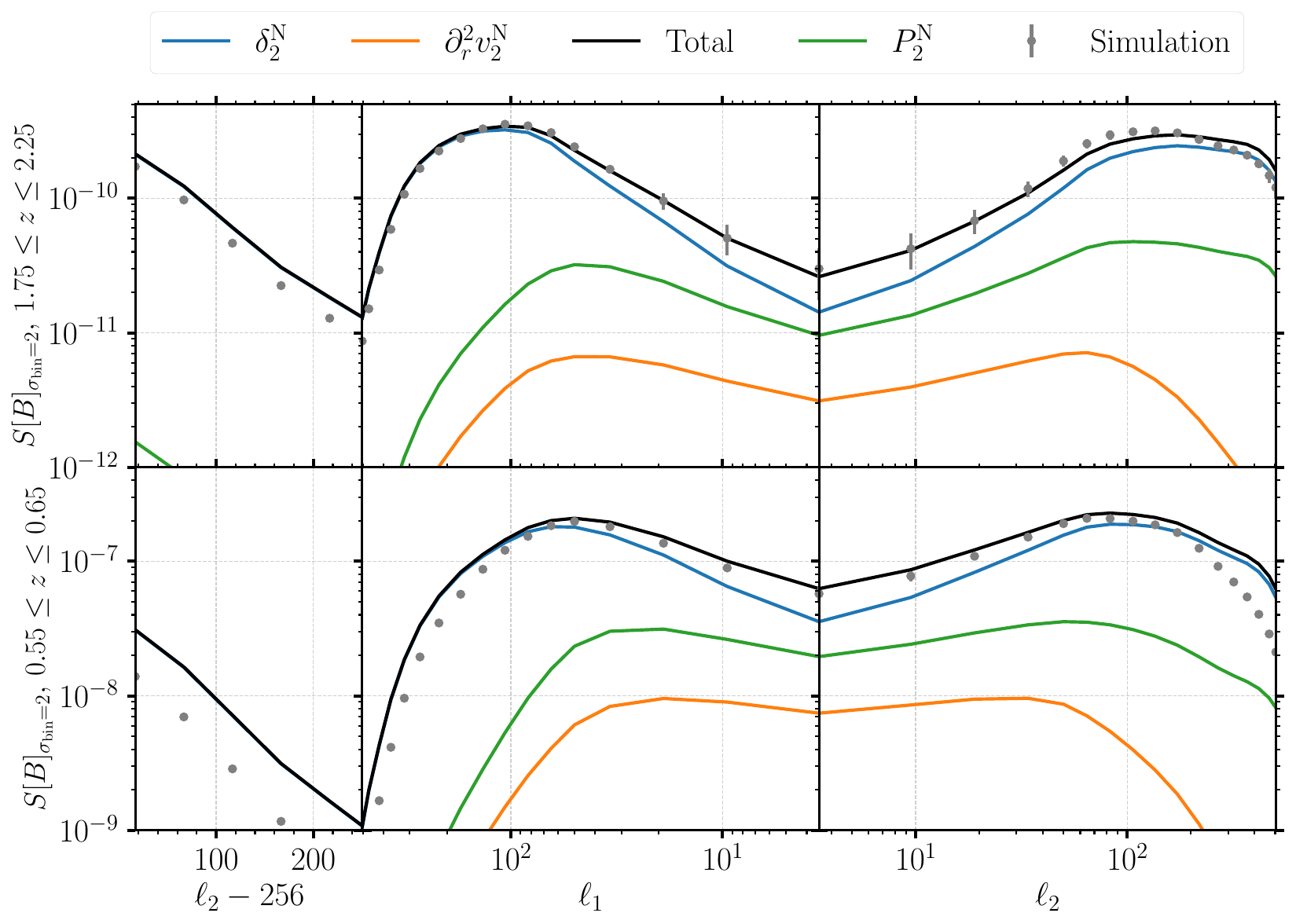}
    \includegraphics[scale=0.48]{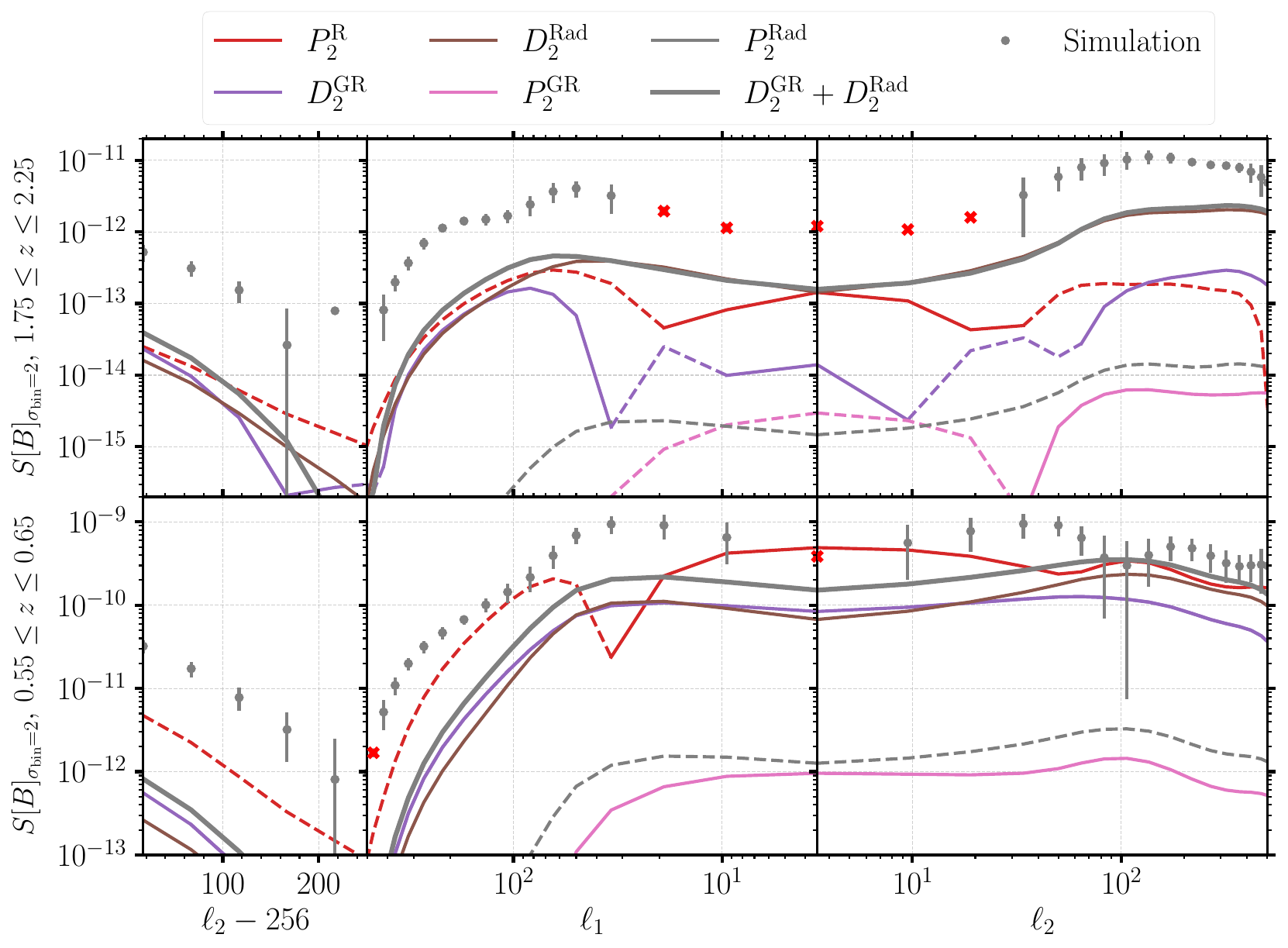}
    \caption{Similar plot as shown in Figs.\,\ref{fig:simVSth} -- \ref{fig:simVSth_rela}, but with a smoothing standard deviation $\sigma_{\rm bin} = 2$. We can see how the smoothing affects the low-redshift total bispectrum by shifting down the power of the measured bispectrum. Similarly, the smoothing increases the power of the measured bispectrum for the relativistic part. }
    \label{fig:smooth2}
\end{figure}

\paragraph{\bf Total bispectrum}
In Fig.\,\ref{fig:simVSth}, we show the binned total bispectrum  (in black) which includes all considered terms as dissected in Eq.\,\eqref{eq:fullbl} except the three last potential terms. In colours, we only show the dominant Newtonian contributions. The grey points labelled with `$\rm Simulation$' are obtained by measuring the bispectrum in the relativistic simulations and setting the potential to zero for the ray tracing algorithm. The error bars indicate the standard error of the mean over the ten paired simulations.
We obtain a quite good agreement between the theory and the simulation except at small scales. This was already observed in the power spectrum analysis of~\cite{Montandon:2022ulz}, and is due to the finite resolution of the simulations. Note that this explains why the discrepancy is stronger at lower redshift since each fixed multipole corresponds to a smaller scale. For the highest redshift bin, we observe good agreement for all configurations except for the smallest scales probed, while for the redshift bin $0.55 \leq z \leq 0.65$, the measurements start to lose power with respect to the theory above $\ell \sim  100$. In the first Fig.\,\ref{fig:smooth2}, we show the results after applying smoothing with $\sigma_{\rm bin} = 2$. At high redshift we observe excellent agreement, while at low redshift a small offset appears, with the simulations showing slightly less power even on large scales. This offset may arise from resolution effects, but could also be due to the smoothing procedure, which we discuss in more detail below.

\paragraph{\bf Relativistic effects}
In Fig.\,\ref{fig:simVSth_rela}, we show all the relativistic contributions to the bispectrum, as well as the measurements (in grey dots). Note that the red plot markers come with error bars that include zero (which we do not show to not obscure the plot), indicating no detection. The bispectrum is smoothed with $\sigma_{\rm bin}=1$, unlike in Fig.\,\ref{fig:simVSth}. In Fig.\,\ref{fig:smooth2}, we can also see the same results but with $\sigma_{\rm bin} = 2$. Before comparing, we can note that for the redshift bin $1.75 \leq z \leq 2.25$, the smoothing of the bispectrum tends to suppress the contribution of the GR part in violet ($D^{\rm GR}_2$) because it oscillates around zero. Hence, for that redshift, almost all dynamical effects are due to radiation. Otherwise, we recover the same kind of result as in Fig.\,\ref{fig:term_comparison}. At lower redshift, both dynamical effects are always positive, avoiding any suppression. The sum of the two effects is then only about a factor $2$ smaller than the relativistic projection effect.

The measurements are obtained by subtracting the bispectra of the relativistic and Newtonian simulations. For the redshift $1.75 \leq z \leq 2.25$, most large-scale triangles yield no detection. In the squeezed limit, we have a significant detection of the bispectrum in the simulations. For the low redshift $0.55 \leq z \leq 0.65$, almost all configurations are significantly detected.

The correct interpretation of these simulation measurements is tricky. Indeed, the subtraction incorporates several numerical artefacts that arise from the standard procedures used to include radiation and relativistic projection effects in Newtonian simulations. Specifically, two additional contributions must be considered in such a measurement:
\begin{itemize}
\item \textbf{Residual gauge effects:} The relativistic ray tracer interprets particle positions in the Poisson gauge, even though in Newtonian simulations they are formally defined in the so-called $N$-body gauge \citep{Fidler:2015npa}. This mismatch has been estimated for the power spectrum by~\cite{Montandon:2022ulz}, where it was found to be subdominant relative to nonlinear relativistic corrections. Its impact on the bispectrum, however, has not been assessed before. 
\item \textbf{Residual backscaling effects:} In Newtonian simulations, radiation is typically included by modifying the initial conditions so that the resulting power spectrum matches the correct radiation-included power spectrum at redshift $z=0$, while evolving only under matter gravitational dynamics. For the power spectrum, this effect has been estimated by~\cite{Brandbyge:2016raj} and \cite{Adamek:2017grt}, and it was found to be $\sim 1\%$ on the largest scales of the matter power spectrum at low redshifts. We are not aware of a similar study for the bispectrum.
\end{itemize}
Therefore, the ``pure'' relativistic bispectrum obtained by~\cite{Montandon:2022ulz} should be understood as the discrepancy between standard Newtonian simulations (employing backscaling) and relativistic simulations including the evolution of radiation.
This discrepancy includes physical contributions --- namely the GR and radiation terms, $B^{D_2^{\rm GR}}_{\ell_1\ell_2\ell_3} + B^{D_2^{\rm Rad}}_{\ell_1\ell_2\ell_3}$ defined in Eq.\,\eqref{eq:b_split} --- and numerical artefacts (residual gauge and backscaling effects). Based on the negligible magnitude of numerical artefacts at the linear level, \cite{Montandon:2022ulz} concluded that their measurements were likely dominated by dynamical relativistic effects.

We are now able to compute independently the dynamical relativistic and radiation effects. Their sum is shown in grey line in Fig.~\ref{fig:simVSth_rela}. As we can see, it is smaller by a factor $\sim 5$ than what was measured in the simulations of~\cite{Montandon:2022ulz}. Hence, we find that numerical errors are comparable to the physical effects. This indicates that in the bispectrum, numerical artefacts can be as significant as the dynamical relativistic contributions. Note that the discrepancy decreases for the lowest redshift bin $0.55 \leq z \leq 0.65$, where the theory agrees better in the squeezed triangle plot and in the large-scale equilateral configuration. This can be explained by the numerical artefacts, since backscaling and gauge effects should decrease with the redshift.

Finally, as already mentioned, we have not included numerical resolution effects that cause a loss of power on small scales. The smoothing can induce a leakage of this loss of power at large scales and an overall shift of the total bispectrum. This is, in addition to resolution effects, one of the possible causes of the offset observed at low redshifts in Fig.\,\ref{fig:smooth2}. The relativistic effects, however, are computed by subtracting two simulations. The leading-order effect causing the small-scale loss of power cancels, leaving only higher-order effects that tend to a constant power. This has been already observed by~\cite{Montandon:2022ulz} for the angular power spectrum of the number count perturbation. However, at small equilateral scales, the theoretical bispectrum decays. This situation could lead to more small-scale power in the relativistic bispectrum as compared to the theory. We conclude that the smoothing should induce a global positive shift at large-scale equilateral and squeezed configurations of the simulation measurements. In the second Fig.\,\ref{fig:smooth2}, we can see the effect of increasing the smoothing scale $\sigma_{\rm bin}$. A larger $\sigma_{\rm bin}$ improves the visibility of the pure relativistic part of the bispectrum, making its shape more evident. In this regime, the analytical bispectrum closely resembles the measured one, though the discrepancy compared to Fig.\,\ref{fig:simVSth_rela} becomes more pronounced. This could therefore explain part of the disagreement we observe in Fig.\,\ref{fig:simVSth_rela}.

To conclude, we compared our theoretical computation and numerical implementation with the simulation measurements of~\cite{Montandon:2022ulz}. The dominant Newtonian part agrees very well with the simulation, except at small scales and low redshift, where numerical effects start to become important in the simulation. Moreover, at small scales, nonlinear effects become more important, and the tree-level bispectrum is no longer an accurate description of the underlying physical processes. Thanks to the analytical calculation, we can see the contribution of each term in the number count perturbation. For the pure relativistic part, we have recovered the right shape and, at low redshift, compatible measurements. Contrary to the conclusions of~\cite{Montandon:2022ulz}, it seems that the additional numerical error, i.e. the residual gauge and backscaling effects, are comparable in amplitude with the dynamical relativistic and radiation effects. Moreover, to obtain a good detection of the bispectrum, we need to use a smoothing procedure with a large enough scale, at least $\sigma_{\rm bin} = 1$ for the bin $1.75 \leq z \leq 2.25$. But this method mixes different scales so that numerical resolution effects that are usually limited to small scales can leak to all scales and generate a considerable shift of the power.

On the theory side, we are now able to separate the radiation and GR effects, and compare them to the relativistic projection effects that are much more studied. We can therefore conclude that the dynamical effects are of the same order as the projection effects for the two redshifts and for all the configurations studied. For the redshift bin redshift bin $1.75 \leq z \leq 2.25$, we find that radiation is the main dynamical relativistic effect, dominating even the projection effects by more than one order of magnitude in the squeezed limit. For the redshift bin $0.55 \leq z \leq 0.65$, the sum of GR and radiation effects is only a factor $2$ smaller than the projection effects.

\section{Conclusions}\label{sec:conclusion}
In this article, starting from the first- and second-order number count results \citep{Yoo:2009au,Yoo:2010ni,Challinor:2011bk,Bonvin:2011bg,Jeong:2011as,Yoo:2014sfa,Bertacca:2014dra,DiDio:2014lka,Magi:2022nfy}, we have numerically evaluated the redshift binned angular bispectrum of the galaxy number count without relying on the Limber approximation. Our calculation includes redshift binning and consistently incorporates all Newtonian terms (density, redshift-space distortions, and quadratic terms), non-integrated projection effects, as well as all dynamical relativistic and radiative contributions. 
Integrated terms, such as lensing, can be neglected for auto-bispectra \citep{DiDio:2015bua}, and removed from our simulations measurements.
By extending the computation of \cite{Assassi:2017lea}, we reduced the problem to one-dimensional integrals, making the numerical computation of the considered bispectrum contributions feasible. In total, we have incorporated 7 linear and 14 second-order terms (summarised in Eqs.\,\ref{eq:1stNB} and~\ref{eq:2ndNB} respectively) and evaluated the resulting bispectrum for various triangle configurations. Our main analytical bispectrum results -- presented in a computationally tractable form -- are summarised in Eq.\,\eqref{eq:fullbl}, with the individual contributions given in Eqs.\,\eqref{eq:density_term}--\eqref{eq:davd1v}. 

We make our computations publicly available via the \href{https://github.com/TomaMTD/ang_bispec}{\texttt{ang\_bispec}} code. Its computation time is largely dominated by the evaluation of the generalised power spectra defined in Eqs.\,\eqref{eqs:generalized_ps_def}. Only seven spectra are needed to compute all 14 bispectrum terms. In the most demanding case -- namely, for a triplet of distinct multipoles $\ell$ and redshifts $z$ -- evaluating these seven generalized power spectra requires approximately $10$~CPU hours. Once determined, computing any of the 14 bispectrum terms becomes inexpensive, requiring only about $10$~CPU seconds per term. While this runtime could likely be further optimised, the present implementation is already about one order of magnitude faster than the SFB method of \cite{Benabou:2023ldb}. The difference in runtime is partly due to the distinct applications targeted by the two approaches, and by our integration routines that are accelerated by the efficient evaluation of hyper-geometric functions and FFTLog, first introduced by~\cite{Assassi:2017lea} for the bispectrum.
In the relativistic case, the additional $r_1$-integrals of Eqs.\,\eqref{eq:density_term}-\eqref{eq:projection1} must also be computed, and these have a computational complexity comparable to that of the generalised power spectra.

The computations presented in this work push the range and accuracy of existing state-of-the-art in angular bispectrum predictions \citep{DiDio:2014lka, DiDio:2015bua, Assassi:2017lea, DiDio:2018unb}, which were either limited to the pure density term and/or relied on the Limber approximation for redshift binning and redshift-space distortions. On the numerical side, measurements of the full angular matter bispectrum and its general relativistic components were previously performed by \cite{Montandon:2022ulz}, which our work complements and extends. Another approach, the SFB formalism -- mathematically related to our method and sometimes referred to as tomographic spherical harmonics (TSH) -- was used by \cite{Benabou:2023ldb} to estimate the dominant Newtonian terms beyond the Limber approximation, though without redshift binning. While the two approaches are formally equivalent, their practical implementations differ substantially. In particular, the mathematical relation between them cannot be directly used to translate the TSH bispectrum into the SFB bispectrum. Furthermore, they are aimed at different applications: SFB is more suited to spectroscopic surveys, which benefit from its higher line-of-sight mode resolution, whereas TSH is directly adapted to photometric surveys with a small number of wide redshift bins.

Looking forward, our approach is particularly suited for application to next-generation photometric surveys such as Euclid, which will provide galaxy maps across $\sim 6$ redshift bins covering one third of the sky. While these maps are primarily intended for weak-lensing studies, they are also ideally suited for angular statistics of the galaxy number count. The formalism developed here thus enables a complementary analysis pipeline, bridging between the two main cosmological observables: lensing and galaxy clustering.

Among the Newtonian terms (density, redshift-space distortion and quadratic terms), we found that the density is, as expected, the dominant term for the small-scale configurations, whereas the quadratic terms are dominant in the large-scale equilateral configuration and have a significant contribution in the squeezed limit. By studying each Newtonian quadratic term, we found that the whole term is the result of two cancellations for which we provide a theoretical explanation; see Appendix~\ref{app:cancellation}. We made a comparison with the recent simulation measurements performed by~\cite{Montandon:2022ulz}. We have shown that it matches well our analytical computation, except for small scales/lower redshift where numerical effects start to decrease the power. We also discussed the measurements employing smoothing, which show excellent agreement in the highest redshift bin but exhibit an offset -- even on large scales -- in the lowest redshift bin. Since a given multipole corresponds to a smaller physical scale at low redshift, this offset can be attributed to the limited resolution of the simulations of \cite{Montandon:2022ulz}, which should be rectified in future work. In addition, the smoothing procedure itself may contribute to the offset by mixing different triangle configurations.

We incorporate in our computations also general relativistic and radiative (GR) effects, and perform a comparison with the relativistic effects numerical estimation of \cite{Montandon:2022ulz}. We have evaluated the angular bispectrum for two redshifts bins: $1.75 \leq z \leq 2.25$ and $0.55 \leq z \leq 0.65$. The comparison with the simulation was more challenging than for the Newtonian part. Indeed, the numerical estimate of \cite{Montandon:2022ulz} represents the total discrepancy between standard Newtonian simulations -- i.e., Newtonian gravity supplemented with residual gauge errors from the relativistic ray tracing treatment and backscaling [to account indirectly/passively for radiation \citep{Brandbyge:2016raj, Adamek:2017grt}] -- and a consistent relativistic simulations employing GR gravity, gauge-consistent ray-tracing, and a forwards approach that actively evolve radiation perturbations; see~\cite{Montandon:2022ulz} and Fig.\ 7 of~\cite{Angulo:2021kes}). Comparing this estimate with simulations and with our theoretical evaluation of dynamical relativistic and radiation effects therefore only reveals the fraction of the measured signal that originates from physical effects, with the remainder attributable to residual gauge and backscaling errors.

An additional source of discrepancy may also arise from the smoothing procedure, which 
we deem necessary for achieving a significant detection of such tiny effects in the bispectrum. Smoothing can introduce a systematic bias by leaking small-scale numerical effects into large scales. We have seen that this mechanism can also contribute to the low-redshift offset observed in the total bispectrum measurements, and we therefore suggest that it may account for part of the shift observed in the pure relativistic result (see, e.g., Fig.\,\ref{fig:simVSth_rela}).

We have included leading relativistic projection effects only up to $\mathcal H/k$, as no consensus currently exists for higher-order contributions~\citep{Yoo:2014sfa,Bertacca:2014dra,DiDio:2014lka,Magi:2022nfy}; see the end of the introduction for details. Since we can remove these effects in our simulations, we also limited our analysis to the terms not involving line-of-sight effects and leave them for future work. 
These relativistic projection effects constitute the main focus of the community studying relativistic effects in large-scale structure, since in terms of the relativistic expansion parameters $\mathcal H/k$, they are the dominant ones.
It is therefore important to compute them and compare them to the dynamical GR and radiation effects. For the redshift bin $1.75 \leq z \leq 2.25$, we have seen that these have a similar amplitude as the relativistic projection effects. For large scales, we argue that this is expected since at $z \sim 2$, the cosmological horizon corresponds roughly to a multipole of $1.2$, hence making the expansion in $\mathcal H/ k$ irrelevant. In the squeezed limit, we have seen that the radiation effects are the dominant relativistic effects by at least one order of magnitude. For the redshift bin $0.55 \leq z \leq 0.65$, the scales considered are smaller, which makes projection effects the dominant relativistic effects. However, the sum of the GR and radiation effects is only a factor two smaller.

Finally, we now have access to the contribution of each term in the relativistic bispectrum, and thus have improved control of potential systematic errors. We found that for high redshifts, the pure relativistic part is dominated by the radiation effects because the GR part is suppressed by the smoothing. For low redshifts, the dynamical effects are of the same order of magnitude as the projection effects. Overall, for the redshift and scales studied, we cannot neglect the dynamical effects when studying the projection effects with the number count angular bispectrum. Note also that our computations quantify the orders of magnitudes of the pure relativistic effects compared to the total amplitude.  

To derive our theoretical results, we have made several simplifications in our analysis, detailed and justified at the end of the introduction. In particular, we neglected galaxy bias, evolution bias, and magnification bias; a natural next step would be to address these simplifications which are in principle straightforward following \cite{DiDio:2018zmk}. Including line-of-sight terms following \cite{DiDio:2015bua} and \cite{Simonovic:2017mhp} are also a natural next step to fit observations, however require the computations of more integrals. 
Another avenue for future work involves incorporating primordial non-Gaussianity, which can be readily implemented within our theoretical framework \cite[see e.g.][]{DiDio:2016gpd}. 
Once these obvious limitations are addressed, we will be equipped with a theoretical pipeline that could assist in addressing potential degeneracies between primordial and intrinsic non-Gaussianity, and devise tests for the theory of gravity in the nonlinear sector.

\section*{Acknowledgements}
The numerical results presented here have been achieved using the Vienna Scientific Cluster (VSC) and the Centre de Calcul de l’IN2P3 (CC-IN2P3). We thank Bartjan van Tent for sharing the binned bispectrum estimator code. TM thanks Ruth Durrer, Oliver Hahn and Vivian Poulin for support and many helpful discussions. TM is supported by funding from the European Research Council (ERC) under the European Union’s HORIZON-ERC-2022 (grant agreement no.\ 101076865). ED acknowledges funding from the European Research Council (ERC) under the European Union’s Horizon 2020 research and innovation program (Grant agreement No.~863929; project title “Testing the law of gravity with novel large-scale structure observables”). The work of JA is supported by the Swiss National Science Foundation.

\paragraph{Note added} After the initial version of this paper has been submitted, a related study appeared on arXiv \citep{Villey:2025xfz}, which provides a new calculation of observables up to second order.

\appendix
\counterwithin{figure}{section} 

\section{Fast Fourier Transform in log-space}\label{app:fftlog}

The idea of using FFTLog was introduced in cosmology by \cite{ Simonovic:2017mhp, McEwen:2016fjn, Schmittfull:2016jsw, Hamilton:1999uv}. Thanks to a discrete algorithm, it allows one to expand a given function of $k$, $g(k)$, as a sum of complex power laws. 
In our case, we expand 
\begin{equation}\label{eq:pk_decomp}
    g(k) = \sum_{p=-N/2}^{N/2} c_p k^{b+i \eta_p}\,,\qquad \text{where}\qquad  \eta_p = \frac{2\pi p}{\log{(k_{\rm max}/ k_{\rm min})}}\,,
\end{equation}
and where $N$ is the number of modes used to tabulate $g(k)$. The coefficients $c_p$ take the form 
\begin{equation}
    c_p = \frac{1}{N} \sum_{l=0}^{N-1} g(k_l) k_l^{-b} k_{\rm min}^{-i \eta_p} e^{-\frac{2i\pi p l}{N}} \,.
\end{equation}
Now, thanks to Eq.\,\eqref{eq:math}, we can express the integral of $g(k) j_\ell(k r)j_\ell(k \chi)$ as 
\begin{eqnarray}\label{eq:Cofchi_decom}
\int dk g(k) j_\ell(k r)j_\ell(k \chi) = \frac{\pi}{2r^2} \sum_{p=-N/2}^{N/2} c_p I_\ell(\nu_p, r, \chi) \,.
\end{eqnarray}
In our case, we use Eq.\,\eqref{eq:Cofchi_decom} to evaluate Eqs.\,\eqref{eqs:generalized_ps_def} as well as to calculate the radiation terms which appear in Eqs.\,\eqref{eq:density_term}, \eqref{eq:G2}, and \eqref{eq:projection1} as a sum over $p$. 

\section{Second-order perturbation theory kernels} \label{app:Kernel}
In second-order perturbation theory, fields can be written in Fourier space as a convolution integral such as Eqs.\,\eqref{eq:def_delta2} and \eqref{eq:def_v2} where the kernel can be denotted in general as $\mathcal K_{\rm X}^{\mathcal I}$, where the index $\rm X$ stands for Newtonian (N), general relativistic (GR) or radiation (Rad), and $\mathcal I$ can be $\delta$ or $v$. In this section, we give the expressions of the kernels that are needed for our derivation. Following \cite{Tram:2016cpy}, the kernels defined in Eqs.\,\eqref{eq:def_delta2} and \eqref{eq:def_v2} read \citep{Villa:2015ppa, Tram:2016cpy}
\begin{equation}
    \label{eq:kernel}
      \mathcal K^{\mathcal I}_{\rm X} (k_1,k_2,k_3) =  
       \beta_{\rm X}^{\mathcal I}-\alpha_{\rm X}^{\mathcal I} + \frac{\beta_{\rm X}^{\mathcal I}}{2} \mu \left(\frac{k_2}{k_3}+\frac{k_3}{k_2}\right) + \alpha_{\rm X}^{\mathcal I} \mu^2 + \gamma_{\rm X}^{\mathcal I}\left(\frac{k_2}{k_3}-\frac{k_3}{k_2}\right)^2\,,  
\end{equation}
with $\mu = \boldsymbol{k}_2 \cdot \boldsymbol{k}_3 / ( k_2 k_3)$. 
In the case of the density, the Newtonian coefficients are given by
\begin{equation}\label{eq:VR_N}
    \alpha^{\delta}_{\rm N} = \frac{7-3v}{14}, \quad 
    \beta^{\delta}_{\rm N} = 1, \quad 
    \gamma^{\delta}_{\rm N} = 0 \,.
\end{equation}
The GR terms are given by 
\begin{equation}
    \label{eq:VR_GR}
\begin{split}
    \alpha^{\delta}_{\rm GR} =& \left( 4f +\frac{3}{2} \Omega_{m}-\frac{9}{7}w \right)\frac{\mathcal H^2}{k_1^2} + \left( 18f^2+9f^2\Omega_m-\frac{9}{2}f\Omega_m \right)\frac{\mathcal H^4}{k_1^4}\,,\\ 
    \beta^{\delta}_{\rm GR} =& \left( -2f^2 + 6f-\frac{9}{2}\Omega_m \right) \frac{\mathcal H^2}{k_1^2} + \left( 36f^2+18f^2\Omega_m \right)\frac{\mathcal H^4}{k_1^4}\,,\\
    \gamma^{\delta}_{\rm GR} =& \frac{1}{2}\left( -f^2 + f-3 \Omega_m\right) \frac{\mathcal H^2}{k_1^2} + \frac{1}{4} \left( 18f^2+9(f^2-f)\Omega_m \right)\frac{\mathcal H^4}{k_1^4}\,.
\end{split}
\end{equation}
Finally, the radiation terms read 
\begin{equation} 
    \alpha^{\delta}_{\rm Rad} = 0, \quad 
    \beta^{\delta}_{\rm Rad} = 0, \quad 
    \gamma^{\delta}_{\rm Rad} =  -\frac{1}{2} \left( f + \frac{3 \Omega_m}{2} \right) \left(\frac{\mathcal H^2}{k_1^2}+3 f\frac{\mathcal H^4}{k_1^4}\right) \frac{\partial\log{T_\phi}}{\partial\log{k_1}}\,, 
\end{equation}
where $T_{\phi}$ stands for the transfer function of the gravitational potential $\phi$.

The coefficients for the velocity can be obtained by Fourier transform Eq.\,(5.50) of \cite{Villa:2015ppa}. We give here the Fourier transform of each term assuming no primordial non-Gaussianity:
\begin{align}
    - \frac{2 D \dot D}{a H_0^2\Omega_{m0}} \phi_0^2 &\longmapsto -  k_1^2k_2^2 \mathcal N^2 D^2 \frac{9}{2}f \Omega_m \frac{\mathcal H^3}{k^4} \left[4+4\mu^2+4\left( \frac{k_1}{k_2} +\frac{k_2}{k_1} \right)\mu + \left( \frac{k_1}{k_2} -\frac{k_2}{k_1} \right)^2\right]\nonumber\\
    \frac{4 D \dot D}{a H_0^2\Omega_{m0}}\Delta^{-1} \left[ \Delta^{-1} \left(\phi_0^l \phi_0^m\right)_{,lm}-\frac{1}{3} \phi_0^l \phi_{0,l}  \right] &\longmapsto k_1^2k_2^2 \mathcal N^2 D^2 9 f \Omega_m \frac{\mathcal H^3}{k^4} \left[1+\frac{1}{3}\mu^2 + \frac{2}{3}\left( \frac{k_1}{k_2} +\frac{k_2}{k_1} \right)\mu\right] \nonumber\\
    -\frac{4D\dot D}{9(H_0^2\Omega_{m0})^2} \phi_{0,l}\phi_{0}^{,l}
    &\longmapsto k_1^2k_2^2 \mathcal N^2 D^2 f  \frac{\mathcal H}{k^2} \left[ 2\mu^2 + \left( \frac{k_1}{k_2} +\frac{k_2}{k_1} \right)\mu \right]\nonumber\\
    \frac{4\dot F}{9(H_0^2\Omega_{m0})^2} \Delta^{-1} \left[ \phi_{0,lm} \phi_0^{lm}-(\Delta \phi_0)^2  \right]
    &\longmapsto k_1^2k_2^2 \mathcal N^2 D^2 \frac{6 w }{7}  \frac{\mathcal H}{k^2} \left[ 1-\mu^2\right]
\end{align}
Noting the different convention of the second-order quantities $v_2 = v_2^{\rm VR} /2$ where $v_2^{\rm VR}$ refers to the second-order velocity used by~\cite{Villa:2015ppa}, we can compute the factors 
\begin{align} \label{eq:VRG}
    \alpha^{v}_{\rm N} &= f - \frac{3}{7} w ,   &
    \beta^{v}_{\rm N} &= f ,&
    \gamma^{v}_{\rm N} &= 0 
    \,,\\
    \alpha^{v}_{\rm GR} &= - \frac{15}{2} \Omega_m f \frac{\mathcal H^2}{k_1^2}, &  
    \beta^{v}_{\rm GR} &= - 12 \Omega_m f \frac{\mathcal H^2}{k_1^2}, &
    \gamma^{v}_{\rm GR} &= -\frac{9}{4} \Omega_m f \frac{\mathcal H^2}{k_1^2}\
    \,,
\end{align}
while for radiation we have $\alpha^{v}_{\rm Rad} = \beta^{v}_{\rm Rad} = 0$ and 
\begin{equation}
    \label{eq:VRG_Rad}
\begin{split}
    \gamma^{v}_{\rm Rad} =& -\frac{1}{2} \frac{1}{1+3\frac{\mathcal H^2}{k_1^2}}\left( f + \frac{3 \Omega_m}{2} \right) \left(\frac{\mathcal H^2}{k_1^2}+3 f\frac{\mathcal H^4}{k_1^4}\right) \frac{\partial\log{T_\phi}}{\partial\log{k_1}}\\
    \simeq& - \frac{1}{2} \left( f + \frac{3 \Omega_m}{2} \right) \left(\frac{\mathcal H^2}{k_1^2}+3 (f-1)\frac{\mathcal H^4}{k_1^4}\right)   \frac{\partial\log{T_\phi}}{\partial\log{k_1}}
    \,,
\end{split}
\end{equation}
which is obtained by injecting the radiation term of the density in Eqs.~(2.21) and (2.19) of~\cite{Adamek:2021rot}. 

To arrive at a compact form, it is convenient to expand the coefficient in a power series of $\mathcal H / k$, e.g.\ the total coefficient without radiation $\alpha^{\mathcal I} = \alpha^{\mathcal I}_{\rm N} + \alpha^{\mathcal I}_{\rm GR}$ can be written as $\alpha = \sum_i \alpha_i (\mathcal H / k)^{2i}$, and similarly for $\beta^{\mathcal I}$ and $\gamma^{\mathcal I}$. Using these variables, and ignoring the radiation term 
for the moment, we can decompose the total kernels $\mathcal K^{\mathcal I} = \mathcal K^{\mathcal I}_{\rm N} + \mathcal K^{\mathcal I}_{\rm GR} + \mathcal K^{\mathcal I}_{\rm Rad}$ in powers of $k_1$, see Eq.\,\eqref{eq:kernel_decomp}, and find that the non-vanishing coefficients $f_{mn}$ are given by
\begin{align}\label{eqs:fnl_coef}
    \frac{f^{(-4)}_{00}}{\mathcal H^4} &= \frac{\beta_2 - \alpha_2}{2} - 2\gamma_2\,,
    & \frac{f^{(-4)}_{2,-2}}{\mathcal H^4} &= \frac{\alpha_2-\beta_2}{4}+\gamma_2\,,
    & \frac{f^{(-2)}_{0,-2}}{\mathcal H^2} &= \frac{\mathcal H^2}{2} \left( \frac{\beta_2}{2}-\alpha_2 \right)\,,\nonumber\\
    \frac{f^{(-2)}_{00}}{\mathcal H^2} &= \frac{\beta_1 - \alpha_1}{2}-2\gamma_1\,,
    & \frac{f^{(-2)}_{2,-2}}{\mathcal H^2} &= \frac{\alpha_1-\beta_1}{4}+\gamma_1\,,
    & f^{(0)}_{0,-2} &= \frac{\mathcal H^2}{2} \left( \frac{\beta_1}{2}-\alpha_1 \right)\,,\nonumber\\
    f^{(0)}_{00} &= \frac{\beta_0 - \alpha_0}{2}\,,
    & f^{(0)}_{2,-2} &= \frac{\alpha_0-\beta_0}{4}\,,
    & f^{(2)}_{0,-2} &= \frac{1}{2} \left( \frac{\beta_0}{2}-\alpha_0 \right)\,,\nonumber\\
    f^{(0)}_{-2,-2} &= \frac{\alpha_2 \mathcal H^4}{4}\,,
    & f^{(2)}_{-2,-2} &= \frac{\alpha_1 \mathcal H^2}{4}\,,
    &  f^{(4)}_{-2,-2} &= \frac{\alpha_0}{4}\,.
\end{align}
For the radiation part $\gamma_{\rm Rad}$, we can use the same kind of decomposition for the factor in front of $\partial \log T_\phi / \partial \log k_1$: $\gamma_{\rm Rad } = \sum_i \gamma_i^{\rm Rad} (\mathcal H / k)^{2i}$. We use FFTLog to decompose $\partial \log T_\phi / \partial \log k_1$ such that 
\begin{align}
 f_{00}^{(-2, R)} = -2\gamma^{\rm Rad}_1\mathcal H^2\,, \qquad &f_{2, -2}^{(-2, R)} = \gamma^{\rm Rad}_1 \mathcal H^2\,, \nonumber\\
 f_{00}^{(-4, R)} = -2^{\rm Rad}_2\mathcal H^4\,, \qquad& f_{2, -2}^{(-4, R)} = \gamma^{\rm Rad}_2\mathcal H^4 \,.
\end{align}

\section{Bispectrum resulting from quadratic terms}\label{app:Quadratic terms}
Quadratic terms are second-order terms made from the product of two linear fields. In harmonic space, we have shown that all quadratic terms can be written in the form of Eq.\,\eqref{eq:quadratic} that we re-write here for convenience as
\begin{align}\label{eq:quadratic_bis}
   b^{XY}_{\ell_1 \ell_2 \ell_3} 
    &=  \int dr_1   \tilde W_{r_1} C^{X}_{\ell_2}(r_1) C^{Y}_{\ell_3}(r_1) + 5\times \circlearrowleft\,.
\end{align}
In the following, we give explicitly the expressions of all $10$ bispectra hidden in this compact form and give some theoretical hints to explain the cancellations that we have phenomenologically observed between the Newtonian terms.  

\subsection{Cancellations between quadratic terms} \label{app:cancellation}

From Fig.\,\ref{fig:quadratic}, we observe significant cancellations among some quadratic terms. These cancellations can be understood by examining the second-order number count. In particular, summing these pairs of terms yields to
\begin{eqnarray}
\label{eq:cancellation1}
    \left( \partial_r^2 v_1 \right)^2 + \partial_r v_1 \partial_r^3 v_1 &=& \frac{1}{2} \partial_r^2 \left( \partial_r v_1 \right)^2 \ ,
    \\
    \label{eq:cancellation2}
    \partial_r v_1 \partial_r \delta_1 + \partial_r^2 v_1 \delta_1 &=& \partial_r \left( \partial_r v_1 \delta_1 \right) \, .
\end{eqnarray}
By summing these terms, we find that they can be expressed as global radial derivatives of the squares of linear perturbations. Therefore, their contribution to the bispectrum is dominated by radial modes. However, when binning in redshift (particularly in the case of photometric redshift bins), these radial modes are strongly suppressed, resulting in the cancellations as observed in Fig.\,\ref{fig:quadratic}. To compute the bispectrum contribution of the terms on the left-hand side of Eqs.\,(\ref{eq:cancellation1}) and~(\ref{eq:cancellation2}), we perform one (or two) integrations by parts over $r_1$ , reducing the terms to lower-order contributions in the weak-field expansion. The amplitude of these lower-order terms effectively quantifies the accuracy of the cancellations among the quadratic terms.

Such cancellations can be computed explicitly. By starting from the two terms 
\begin{align}
 b^{(\partial_r^2 v_1)^2}_{\ell_1 \ell_2 \ell_3} 
   & = 2 
     \int dr_1   \tilde W_{r_1} f^2_{r_1} C^{(0, 2)}_{\ell_2}(r_1) C^{(0, 2)}_{\ell_3}(r_1) + 3\times \circlearrowleft\, , \\
  b^{\partial_r v_1\partial_r^3 v_1}_{\ell_1 \ell_2 \ell_3} 
    &= 
     \int dr_1 \tilde W_{r_1} f^2_{r_1} C^{(1, 3)}_{\ell_2}(r_1) C^{(-1,1)}_{\ell_3}(r_1) + 5\times \circlearrowleft
\,.
\end{align}
We remark that in the generalised power spectrum definition, Eq.\,\eqref{eq:def_gen_spectrum}, the first superscript denotes the order in the weak-field expansion, while the second denotes the number of derivatives on the spherical Bessel functions. As expected, the two terms above are zeroth-order in the weak-field expansion, and therefore are considered Newtonian contributions.
To understand the strong cancellation between these two terms we compute
\be
    k_2^3 k_3^3\frac{d^2}{dr_1^2} \left[ j'_{\ell_2} \left( k_2 r_1 \right) j'_{\ell_3} \left( k_3 r_1 \right) \right] 
    = 2 k_2^4 k_3^4 j''_{\ell_2} \left( k_2 r_1 \right) j''_{\ell_3} \left( k_3 r_1 \right)  + k_2^5 k_3^3 j'''_{\ell_2} \left( k_2 r_1 \right) j'_{\ell_3} \left( k_3 r_1 \right) +k_2^3 k_3^5 j'_{\ell_2} \left( k_2 r_1 \right) j'''_{\ell_3} \left( k_3 r_1 \right) \,. 
\ee
Then by integrating by parts twice on $r_1$ we obtain
\begin{equation}
     b^{(\partial_r^2 v_1)^2}_{\ell_1 \ell_2 \ell_3}  + b^{\partial_r v_1\partial_r^3 v_1}_{\ell_1 \ell_2 \ell_3} = 
     \int dr_1   \frac{d^2}{dr_1^2} \left[\tilde W_{r_1} f^2_{r_1} D_{r_1}^2 \right] D_{r_1}^{-2} C^{(-1, 1)}_{\ell_2}(r_1) C^{(-1, 1)}_{\ell_3}(r_1) + 3\times \circlearrowleft \, .
\end{equation}
We notice that the sum of the two bispectra is given by two generalised spectra, each of them first order in the weak-field expansion. Therefore we do expect that the contribution of these two terms together is much more suppressed than all the other individual Newtonian contributions in the sub-horizon regime $\HH \ll k$, if the second derivative of the window function does not lead to any enhancement. 
At this point, it remains to study the behavior of the following two terms
\begin{align}
   b^{\partial_r v_1\partial_r \delta_1}_{\ell_1 \ell_2 \ell_3} = 
     \int dr_1   \tilde W_{r_1} f_{r_1} C^{(-1, 1)}_{\ell_2}(r_1) \left( C^{(1, 1)}_{\ell_3}(r_1) + 3 f\left( r_1  \right) \HH^2\left( r_1 \right) C_{\ell_3}^{(-1,1)} \left( r_1 \right) \right) + 5\times \circlearrowleft \,, \\
b^{\partial_r^2 v_1 \delta_1}_{\ell_1 \ell_2 \ell_3}  = 
     \int dr_1   \tilde W_{r_1} f_{r_1} C^{( 0,2)}_{\ell_2}(r_1) \left( C^{(0, 0)}_{\ell_3}(r_1) + 3 f\left( r_1  \right) \HH^2\left( r_1 \right) C_{\ell_3}^{(-2,0)} \left( r_1 \right) \right) + 5\times \circlearrowleft\,.
\end{align}
Using
\begin{equation}
    k_2^3 k_3^4 \frac{d}{dr_1} \left[ j'_{\ell_2} \left( k_2 r_1 \right) j_{\ell_3} \left( k_3 r_1 \right) \right] 
= k_2^3 k_3^5 j'_{\ell_2} \left( k_2 r_1 \right) j'_{\ell_3} \left( k_3 r_1 \right) +  k_2^4 k_3^4 j''_{\ell_2} \left( k_2 r_1 \right) j_{\ell_3} \left( k_3 r_1 \right)  
\end{equation}
and
\begin{equation}
k_2^3 k_3^2 \frac{d}{dr_1} \left[ j'_{\ell_2} \left( k_2 r_1 \right) j_{\ell_3} \left( k_3 r_1 \right) \right] 
= k_2^3 k_3^3 j'_{\ell_2} \left( k_2 r_1 \right) j'_{\ell_3} \left( k_3 r_1 \right) +  k_2^4 k_3^2 j''_{\ell_2} \left( k_2 r_1 \right) j_{\ell_3} \left( k_3 r_1 \right)  \, ,
\end{equation}
we find
\begin{eqnarray}   
 b^{\partial_r v_1\partial_r \delta_1}_{\ell_1 \ell_2 \ell_3} + b^{\partial_r^2 v_1 \delta_1}_{\ell_1 \ell_2 \ell_3} &=&
 - \int dr_1 \frac{d}{d r_1}  \left[\tilde W_{r_1} f_{r_1} D_{r_1}^2 \right] D_{r_1}^{-2}
 C^{(-1,1)}_{\ell_2} \left( r_1 \right)  C^{(0,0)}_{\ell_3} \left( r_1 \right)  \nonumber \\
&&
 - 3 \int dr_1 \frac{d}{d r_1}  \left[\tilde W_{r_1} f^3_{r_1} \HH^2_{r_1}D_{r_1}^2 \right] D_{r_1}^{-2}
 C^{(-1,1)}_{\ell_2} \left( r_1 \right)  C^{(-2,0)}_{\ell_3
} \left( r_1 \right) + 5\times \circlearrowleft\, . \qquad
\end{eqnarray}
We see that that the sum of these two terms is given by two contributions: a contribution that is first order in the weak-field expansion, and a contribution that is third order (which is due to the GR contribution to the density fluctuation of the order $\Hc^2/k^2$). Therefore, we expect a weaker cancellation of these two terms with respect to the ones above that involve only the velocity potential.

From Fig.\,\ref{fig:quadratic}, we indeed see a stronger cancellation between the first pair of terms, involving only the velocity potential, than the second pair, as expected by the power counting of the weak-field expansion parameter. 
Numerically we observe that $b_{\ell_1 \ell_2 \ell_3}^{\partial_r^2 v \delta} \simeq -2 b_{\ell_1 \ell_2 \ell_3}^{\partial_r v \partial_r \delta}$
leading to
\begin{equation}
    b_{\ell_1 \ell_2 \ell_3}^{\partial_r^2 v \delta} + b_{\ell_1 \ell_2 \ell_3}^{\partial_r v \partial_r \delta} \simeq -b_{\ell_1 \ell_2 \ell_3}^{\partial_r v \partial_r \delta}
\end{equation}
and, neglecting the GR contributions,
\begin{equation}    
 \int dr_1   \tilde W_{r_1} f_{r_1} C^{(-1, 1)}_{\ell_2}(r_1)  C^{(1, 1)}_{\ell_3}(r_1)   \simeq 
 \int dr_1 \frac{d}{d r_1}  \left[\tilde W_{r_1} f_{r_1} D_{r_1}^2 \right] D_{r_1}^{-2}
 C^{(-1,1)}_{\ell_2} \left( r_1 \right)  C^{(0,0)}_{\ell_3} \left( r_1 \right) \, .
 \end{equation}
Since, in our counting scheme, we would have expected the term on the left-hand side to be one order larger than the one on the right-hand side, we suspect that the radial derivative of the sharp top-hat window function may enhance the integral of the right-hand side.

\begin{figure}
    \centering
    \includegraphics[scale=0.5]{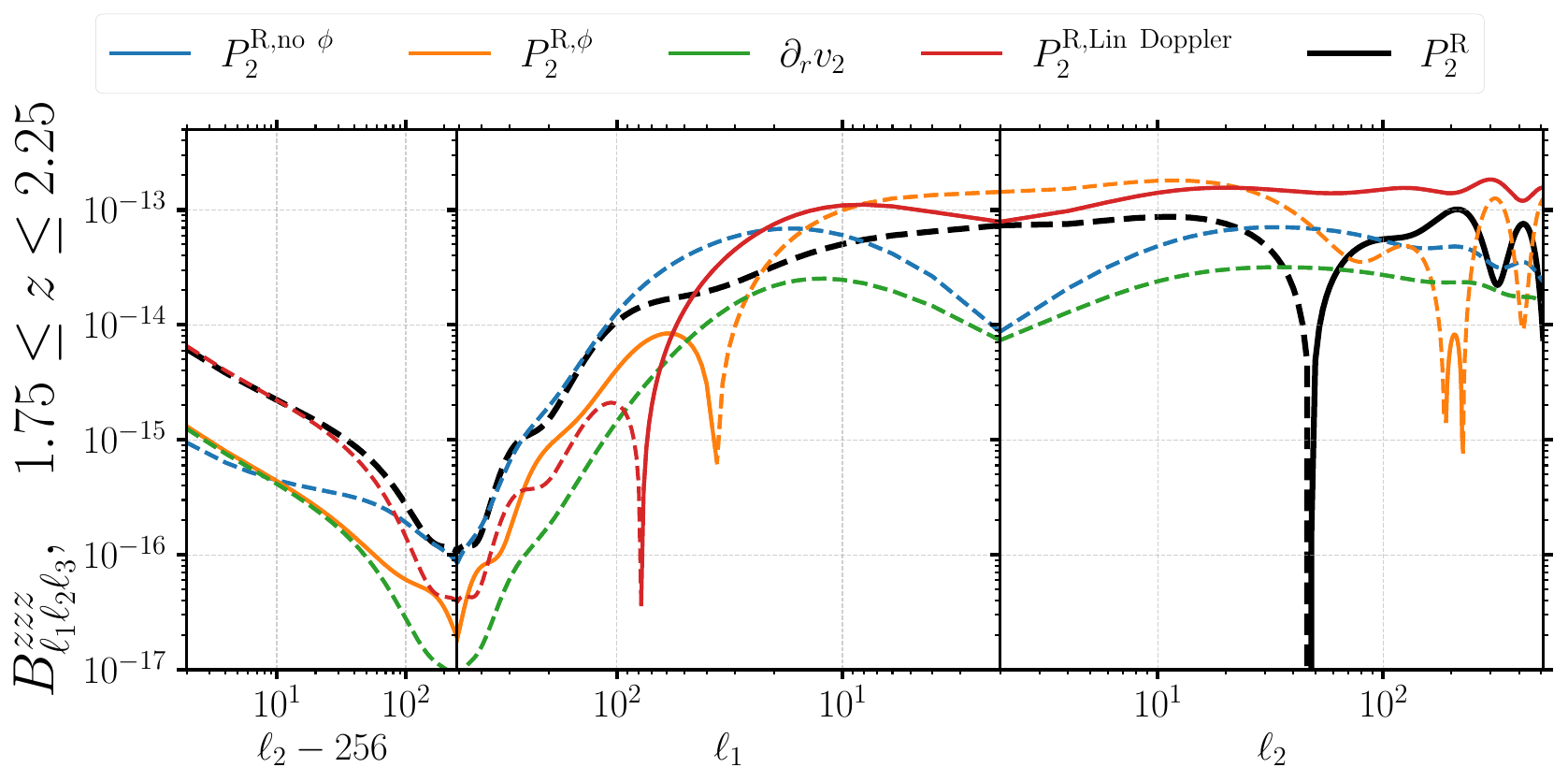}
    \caption{We show each contribution to the Newtonian quadratic term $P^{\rm N}_2$ according to \eqref{eq:QN} for the redshift bin $1.75 \leq z \leq 2.25$. The axis are the same as Fig.\,\ref{fig:term_comparison}.}
    \label{fig:quadraticPR}
\end{figure}

\subsection{Projection effects (first order weak-field expansion)}\label{app:PR}
The quadratic terms read 
\begin{align}\label{eqs:nopot}
    b^{\partial_r v \partial^2_rv}_{\ell_1 \ell_2 \ell_3} &= -
     \int dr_1  \tilde W_{r_1} D^2_{r_1} f^2_{r_1} \mathcal H_{r_1} \left( 1+3 \frac{\dot {\mathcal H}_{r_1}}{\mathcal H_{r_1}^2} + \frac{4}{\mathcal H_{r_1} r_1}\right) C_{\ell_2}^{(-2, 1)}\left( r_1 \right) C_{\ell_3}^{(-2,2)}\left( r_1 \right)
     + 5\times \circlearrowleft\,,\nonumber\\
            b^{\partial_r v \delta}_{\ell_1 \ell_2 \ell_3} &= \int dr_1   \tilde W_{r_1} D^2_{r_1}f_{r_1} \mathcal R_{r_1}  C_{\ell_2}^{(-2,1)}\left( r_1 \right) \left(C_{\ell_3}^{(0, 0)}\left( r_1 \right) + 3 \mathcal H_{r_1}^2 f_{r_1}  C_{\ell_3}^{(-2, 0)}\left( r_1 \right)\right)+ 5\times \circlearrowleft\,,\nonumber\\
    b^{\partial_r v \dot \delta}_{\ell_1 \ell_2 \ell_3} &=
      - \int dr_1  \tilde W_{r_1} D^2_{r_1}f^2_{r_1} \mathcal H_{r_1} C_{\ell_2}^{(-2, 1)}\left( r_1 \right)  
      \left( C^{(0, 0)} \left( r_1 \right)
     + 3 \left( \dot{\mathcal H}_{r_1} + \mathcal H_{r_1}^2 \left( \frac{3}{2} \frac{\Omega_{m,r_1}}{f_{r_1}} - 1\right)\right) C^{(-2, 0)}\left( r_1 \right)\right)
     + 5\times \circlearrowleft\,.
\end{align}
Terms involving the potential read 
\begin{align}\label{eqs:pot_terms}
    b^{\psi \partial_r^3 v}_{\ell_1 \ell_2 \ell_3} &= -\frac{1}{\mathcal N}
     \int dr_1 \tilde W_{r_1} \frac{D^2_{r_1}f_{r_1} }{\mathcal H_{r_1} a_{r_1}}  C_{\ell_2}^{(-2, 3)}\left( r_1 \right)  C_{\ell_3}^{(-2, 0)}\left( r_1 \right)
      + 5\times \circlearrowleft\,,\nonumber\\ 
    b^{\psi \partial_r \delta}_{\ell_1 \ell_2 \ell_3} &=\frac{1}{\mathcal N}
        \int dr_1  \tilde W_{r_1} \frac{D^2_{r_1}}{\mathcal H_{r_1} a_{r_1}} 
        \left(\frac{2}{r_1}C^{(-1, 0)}_{\ell_3}(r_1) +  C^{(0, 1)}_{\ell_3}(r_1)+ 3 \mathcal H_{r_1}^2 f_{r_1} C_{\ell_3}^{(-2, 1)}\left( r_1 \right) \right) C_{\ell_3}^{(-2, 0)}\left( r_1 \right) + 5\times \circlearrowleft\,,\nonumber\\ 
    b^{\partial_r v \partial_r^2 \psi}_{\ell_1 \ell_2 \ell_3} &=  \frac{1}{\mathcal N}
        \int dr_1 \tilde W_{r_1} \frac{D^2_{r_1}f_{r_1}}{\mathcal H_{r_1} a_{r_1}}  C_{\ell_2}^{(-2, 1)}(r_1) C_{\ell_3}^{(-2, 2)}(r_1)
            + 5\times \circlearrowleft\,. 
\end{align}
Finally, the last term, already given in Eq.\,\eqref{eq:davd1v}, involves angular derivatives
\begin{align}
    b^{\partial_a v \partial^a \partial_r v}_{\ell_1 \ell_2 \ell_3} &= - 2 \sqrt{\ell_2(\ell_2+1)}\sqrt{\ell_3(\ell_3+1)} \mathcal A_{\ell_1\ell_2\ell_3}
     \int dr_1  \tilde W_{r_1} D^2_{r_1}f^2_{r_1} \mathcal H_{r_1} C_{\ell_2}^{(-2, 0)}(r_1) C_{\ell_3}^{(-2, 1) }(r_1) 
     + 5\times \circlearrowleft \,,
\end{align}
where the geometric factor $\mathcal A_{\ell_1\ell_2\ell_3}$ is defined in Eq.\,\eqref{eq:Alll}.

In Fig.\,\ref{fig:quadraticPR}, we show the three terms of Eqs.\,\eqref{eqs:nopot} in blue ($P_2^{\rm R, no\ \phi}$) and the three terms of Eqs.\,\eqref{eqs:pot_terms} in orange ($P_2^{\rm R, \phi}$). In addition, we have the pure second-order term in green, whose expression is given in Eq.\,\eqref{eq:projection1} and in red the effect due to the linear Doppler effects which is also considered a relativistic projection effect. The sum of all previous contributions is shown in black.

 \section{Numerical implementation}\label{app:code}
To our knowledge, the latest bispectrum code that has been developed was \texttt{byspectrum}\footnote{\url{https://gitlab.com/montanari/byspectrum}}~\citep{DiDio:2018unb}. However, it is limited to Newtonian terms excluding redshift-space distortion and can handle only infinitesimal redshift bins. Here, we account for finite redshift bins for all terms and have extended the computation to the leading relativistic projection effects and to all GR and radiation effects.  
The code is written in \texttt{Python} and uses the just-in-time compilers \texttt{NUMBA}\footnote{\url{https://numba.pydata.org/}}. We have translated the \texttt{MATHEMATICA} notebook developed by \cite{Assassi:2017lea} to \texttt{Python} which allows us to efficiently evaluate the hypergeometric function ${_2}F_1$ of Eq.\,\eqref{eq:math}. The code is separated into different modules; let us take as an example the first term: Eq~\eqref{eq:density_term}.
\begin{itemize}
    \item Linear cosmology: computes all growth functions $D, f, v$ and $w$ solving the differential equations following \cite{Tram:2016cpy} and \cite{Villa:2015ppa}. 
    It also calls \texttt{CLASS}~\citep{Blas:2011rf}, to evaluate the potential transfer function needed for the radiation term and the potential power spectrum.
    \item Another module takes care of the \texttt{FFTLog} transformations of the potential power spectrum and transfer function.
    \item Now we precompute the generalised power spectra $C_\ell(\chi)$ defined in Eqs.\,\eqref{eqs:generalized_ps_def}. Note that in practice, we only need a few generalised power spectra to obtain all of them. In total, we compute $7$ spectra for all $\ell$ and for a chosen number of $\chi$. 
    This part is the bottleneck of the whole computation.
    \item Then we precompute the integrals over $r_1$ of Eqs.\,\eqref{eq:def_delta2}, \eqref{eq:def_v2}, and \eqref{eq:projection1} for all $\ell$ and for some grid of $r_1$. We do it for all $f_{n, m}$ coefficients. 
    \item Finally, we compute the main integral of Eqs.\,\eqref{eq:density_term}, \eqref{eq:G2}, \eqref{eq:projection1}, \eqref{eq:quadratic}, and \eqref{eq:davd1v} for all possible triplets of $(\ell_1, \ell_2, \ell_3)$. 
\end{itemize}
The quadratic terms are somewhat simpler to evaluate once the generalised power spectra have been precomputed. The code can be found on github: \url{https://github.com/TomaMTD/ang_bispec}. 

To produce the result of this paper and to allow comparisons, we have used the same cosmology as \cite{Montandon:2022ulz} that is, $h=0.67556$, $\Omega_{\rm b}=0.048275$, $\Omega_{\rm cdm}=0.26377$, $\Omega_r=9.16714 \times 10^{-5}$, $A_s =2.215 \times 10^{-9}$ and $n_s = 0.9619$.

\bibliography{ref_inspirehep}{}

\end{document}